\documentclass[twocolumn,prd]{revtex4-1}

\usepackage{graphicx}
\usepackage{caption}
\usepackage{subcaption}

\usepackage{tikz}
\usetikzlibrary{through,calc}

\newcommand{\be}{\begin{equation}}
\newcommand{\ee}{\end{equation}}

\newcommand{\bea}{\begin{eqnarray}}
\newcommand{\eea}{\end{eqnarray}}

\usepackage{grffile}
\graphicspath{{plots/}}

\begin{document}

\title{Gluon Dynamics, Center Symmetry and the deconfinement phase transition in SU(3) pure Yang-Mills theory}

\author{P. J. Silva}
\email{psilva@teor.fis.uc.pt}
\author{O. Oliveira}
\email{orlando@fis.uc.pt}
\affiliation{CFisUC, Department of Physics, University of Coimbra, P-3004 516 Coimbra, Portugal}

%%%%%%%%%%%%%%%%%%%%%%%%%    abstract   %%%%%%%%%%%%%
%.......................................................................................................................................
%.......................................................................................................................................
\begin{abstract}
The correlations between the modulus of the Polyakov loop, its phase $\theta$ and the Landau gauge gluon propagator at finite temperature 
are investigated in connection with the center symmetry for pure Yang-Mills SU(3) theory.
In the deconfined phase, where the center symmetry is spontaneously broken, the phase of the Polyakov loop per 
configuration is close to $\theta = 0$, $\pm \, 2 \pi /3$. We find that the gluon propagator form factors associated 
with $\theta \approx 0$ differs quantitatively and qualitatively from those associated to  $\theta \approx \pm \, 2 \pi /3$. 
This difference between the form factors is a property of the deconfined phase and a sign of the spontaneous breaking of the center symmetry. 
Furthermore, given that this difference vanishes in the confined phase, it can be used as an order parameter associated to the deconfinement transition.
For simulations near the critical temperature $T_c$, the difference between the propagators associated to $\theta \approx 0$ and $\theta \approx \pm \, 2 \pi /3$
allows to classify the configurations as belonging to the confined or deconfined phase. This establishes a selection procedure which has a measurable impact in the 
gluon form factors. Our results also show that the absence of the selection procedure can be erroneously taken as lattice artifacts.
\end{abstract}

\pacs{11.15.Ha,11.10.Wx,14.70.Dj}

\maketitle

%.......................................................................................................................................
%.......................................................................................................................................
\section{Introduction}

The investigation of how the dynamics of QCD  is modified by the
temperature and density has been under intensive study,
motivated mainly by the experimental heavy ion programs running at CERN~\cite{Roland2015} and at RHIC~\cite{Trainor2014}.
From the theoretical side, the understanding of the phase diagram of QCD requires 
the extension of the usual theoretical toolkit to address the properties of strong interacting matter.

The simulations of QCD on a spacetime lattice provides \textit{ab initio} first principles results 
on the non-perturbative regime of hadronic phenomena. Lattice QCD simulations are routinely used to investigate 
the zero temperature and zero density properties of hadronic matter, to tackle the temperature dependence of the thermodynamic properties of hadrons
and to access the thermodynamics of hadronic matter at small densities --- see, for example, \cite{Meyer2015,latt2013gattr} and references therein.

For pure SU(3) Yang-Mills theory and at zero density, lattice QCD simulations have  shown the existence of a first order transition with the gluons becoming deconfined
above the critical temperature $T_c \approx 270$ MeV~\cite{Iwasaki1992,Boyd1996,LuciniTeper2004}.
For temperatures above $T_c$ the gauge bosons behave as massive quasi-particles and it is possible to define a gluon mass.
Lattice simulations show that the gluon mass has a value of about 0.5 GeV for temperatures around $T_c$ and its value increases linearly 
with $T$~\cite{SilvaOliveiraetal2014}. On the other hand, for $T < T_c$ gluons are confined within color singlet states and show up only
as constituents of, for example, mesons or glueballs. If one takes into account the quark degrees of freedom, the picture just described
is essentially unchanged. However, in such case we have a crossover~\cite{Bhattacharya2014,Aoki2006}
instead of a first order transition to the deconfined phase,
and the critical temperature is lowered to $T_c \approx 150$ MeV~\cite{Borsanyi2010,Bazavov2012}.

In what concerns the deconfinement phase transition, its order parameter is the Polyakov loop defined, in the continuum and in the Euclidean space, as
\begin{equation}
   L( \vec{x} ) = \frac{1}{N}\mbox{Tr} \left\{ \mathcal{P} \ \exp\left[ i g \int^{1/T}_0 dx_4 \, A_4(x) \right] \right\} \ ,
   \label{Eq:Polyakov_continuum1}
\end{equation}   
where $\mathcal{P}$ stands for path ordering, $T$ is the temperature and $N = 3$ is the number of colours. Its space averaged value
\begin{equation}
  L = \langle L( \vec{x} ) \rangle \, \propto \, e^{-F_q/T} 
\end{equation}   
is a measure of the free energy of a static quark $F_q$~\cite{Kapusta2011}. 
In the confined phase, i.e. for $T < T_c$, $L = 0$  and the quark free energy is infinite,
suggesting that quarks are confined within hadrons.
On the other hand, above $T_c$ the Polyakov loop is equal to one,  which means that $F_q$ vanishes and quarks behave 
essentially as free particle -- see, for example,~\cite{Lo2013} and references therein.

The Polyakov loop depends directly only on the glue content of the theory but 
it distinguishes if quarks are confined or 
behave as quasi-free particles. In what concerns the glue content of QCD, there isn't such
an analogous operator. To the best knowledge of the authors, there isn't an operator from where one can read about the nature of
the gluons, i.e. if they are confined or behave as quasi-free particles. 
As observed in~\cite{SilvaOliveiraetal2014,Aouane12} the properties of the propagator change dramatically when $T$ crosses the
critical temperature and, at least the so-called gluon electric form factor, can be mapped into a free particle propagator over a limited
range of momenta for temperatures above $T_c$.
On the other hand, the gluon magnetic form factor is clearly not compatible with the usual free particle propagator.
 
 The lattice definition of the Polyakov loop reads
\begin{equation}
   L( \vec{x} ) = \prod^{N_t}_{t=0} \, \mathcal{U}_4(\vec{x},t) \ ,
   \label{Eq:Polyakov_lattice}
\end{equation}   
where $\mathcal{U}_4$ is the time-oriented link, and $L$ has the same definition as in the continuum formulation.

In QCD, like in any gauge theory, the gauge fields belong to the algebra of the gauge group and the fields related by a gauge transformation
\begin{equation}
  A^\prime_\mu(x) =  G(x) A_\mu (x) G^\dagger (x)  - \frac{i}{g} \partial_\mu G(x) \, G^\dagger(x) \ ,
\end{equation}
where $G(x) \in SU(3)$ and $g$ is the coupling constant, are physically indistinguishable. The set of gauge related fields is called a gauge orbit.
Choosing a gauge requires peaking a given configuration from each gauge orbit. The choice of the gauge configurations on each gauge orbit
is a delicate issue not completely resolved in gauge theories; see e.g. \cite{Gribov1978,Vandersickel2012} and references therein.
It is known that this choice of the gauge configuration can change the infrared properties of the 
theory~\cite{Silva2004, Silva2007, Silva2010, Sternbeck2013, Cucchieri1997}.

For the group SU(3) one defines its center group
\begin{displaymath}
 Z_3 = \{ 1, e^{i 2 \pi /3}, e^{-i 2 \pi / 3} \} \ ,
\end{displaymath}
whose elements are such that they commute with all elements of the SU(3) group. 
The elements of $Z_3$, 
associated with global gauge transformations,
divide the group SU(3) into equivalent classes. The gauge group associated with the pure Yang-Mills theory is $SU(3)/Z_3$ and
not  SU(3). In full QCD, the theory is not invariant under the replacement of $q(x) \longrightarrow z \, q(x)$, where $z \in Z_3$ and, therefore,
the gauge group associated with full QCD is SU(3).

The difference in the gauge group associated to the pure SU(3) Yang-Mills theory and full QCD implies, for example, that monopole solutions 
of the classical equations of motion exist only in the pure gauge theory; see, for example,~\cite{MoTsun1993} for further details. 

For pure Yang-Mills theory the global gauge transformations associated with $Z_3$ leave unchanged the Green's function generating functional.
This invariance occurs both for the continuum and the lattice formulation of QCD.
From the elements of $Z_3$, it is possible to build gauge transformations which leave the generating functional
invariant but not the Polyakov loops.

Let us consider the lattice formulation of the pure SU(3) gauge theory. The Wilson action and the measure are invariant under a 
center transformation where the links on a given hyperplane $x_4 = const$ are multiplied by some $z \in Z_3$. 
This type of transformation can be viewed as a singular gauge transformation.
On the other hand, the Polyakov loop changes according to $L(\vec{x}) \rightarrow z \, L(\vec{x})$. 
In the confined phase where $L = 0$, the space averaged Polyakov loop is invariant under such center transformations. However,
above the critical temperature, $L \neq 0$ and $L$ acquires a phase under the center transformation.
Above the critical temperature $L$ is no longer invariant under a center transformation and the center symmetry is said to be spontaneously broken. 
Indeed, the simulations performed in~\cite{Gattringer10,Gattringer11,stokes13,Endrodi14} show that (i) the phase of $L$  takes values which match
essentially those of the $Z_3$ elements, (ii) below $T_c$ the various phases of $L$ are equally populated, (iii) above $T_c$ the various phases of $L$ are not populated likewise,
(iv) above $T_c$ one can identify center domains on the lattice, where the phase of $L( \vec{x} )$ are close to a given $Z_3$ element,
(v) above $T_c$ these center domains define large clusters of $L$ that percolate the lattice volume. 
In~\cite{Asakawa13} it was argued that, above $T_c$, the formation of the center domains can explain certain features of the quark-gluon plasma 
observed experimentally.

From the above considerations it follows that, on the lattice, for temperatures higher than $T_c$ one can 
label a given configuration by the phase of $L$. 
Furthermore, given a particular gauge configuration, the center symmetry allows 
to generate a second configuration which, from the point of view of the sampling is equally probable as the original configuration. Indeed,
although $L$ acquires a different phase, the action is exactly the same for both configurations.

If for $T > T_c$ the gauge configurations can be 
labelled by the phases of $L$, how different are the physical properties
that can be associated with such equally probable configurations? Can we distinguish the various center domains?
If, for example, the thermodynamics associated with the various $Z_3$ related configurations differ, do they lead
to the formation of metastable states? The possibility of having new metastable states, depending on the mean life time of these
states, has the potential to change our view of, for example, the history of the Universe. 

A preliminary study correlating chiral symmetry breaking to the phase of the Polyakov loop was performed in ~\cite{Kovacs08}.
According to this work, in the deconfined phase, the breaking of chiral symmetry is correlated with the phase of $L$. 
If chiral symmetry breaking is sensitive to the phase of the Polyakov loop, are there other properties of QCD which
are also correlated with the phase of the Polyakov loop?

In the present work we try to identify similar effects in pure gauge theory. In particular, we try to correlate the gluon propagator,
computed using lattice QCD simulations, with the phase of the Polyakov loop. 
Our results show that, for temperatures above $T_c$, the gluon propagator is quantitatively and qualitatively different for the 
various $Z_3$ related configurations. Furthermore, we find that the propagator associated with the $e^{\pm i 2 \pi /3}$ phases
for the Polyakov loop are indistinguishable, within our statistical precision.
We also observe that the values of the phase of the Polyakov loop are correlated with the infrared behaviour of the longitudinal gluon propagator, 
in particular with its value at zero momentum. Moreover, this correlations can be used to identify the deconfinement phase transition relying only 
on the gluon propagator.  Preliminary results of our work can be found in~\cite{OliveiraSilva2014}.

In the literature there are several studies of the lattice Landau gauge gluon propagator at finite temperature
\cite{Cucchieri07,Fishcer10,Cucchieri10,Cucchieri10b,Bornyakov11,Bornyakov12,Aouane12,Maas12,
Cucchieri12,Oliveira12,Oliveira12b,Oliveira12c,Maas13,SilvaOliveiraetal2014,Mendes14}, both for SU(2) and
SU(3) gauge theories. However, to the best knowledge of the authors, no one has investigated the correlations of the propagator
with the phase of the Polyakov loop. 
Continuous methods have also been applied to the study of the temperature dependence of the Landau gauge 
gluon propagator, see~\cite{Reinosa2014,Reinosa2015,Reinosa2015a,Reinosa2015b,Canfora2015} and references therein, but again
the correlation with the phase of Polyakov loop was not taken into account.

This paper is organised as follows. In Sec. \ref{sec2} the lattice setup, the computation of the gluon field and of the gluon propagator are discussed.
Furthermore, in this section the $Z_3$ sectors are introduced and the dependence of the gluon propagator with the phase of the Polyakov loop
is reported at a temperature well above the critical temperature. 
In Sec. \ref{sec3} we discuss the simulations close to the critical temperature and identify a criterion to determine the phase (confined or deconfined)
of a given configuration in a Monte Carlo simulation.
In Sec. \ref{sec4} the behaviour of the gluon propagator near $T_c$ is investigated, together with a brief discussion of the continuum limit. 
Finally, in Sec. \ref{sec5} we resume and conclude.

%====================================================================
%====================================================================
\section{Lattice Setup, Gluon Propagator, Center symmetry and $Z_3$ sectors \label{sec2}}

In the present work one considers various lattice simulations using the Wilson gauge action for the 
gauge group SU(3) and for different lattice spacings, i.e. $\beta$ values. 
The physical scale used to convert into physical units was taken from the string tension,
following the procedure described in~\cite{SilvaOliveiraetal2014}.

%+++++++++++++++++++++++++++++++++++++++++++++++++++++++++++++++++++++++++++
%+++++++++++++++++++++++++++++++++++++++++++++++++++++++++++++++++++++++++++
%+++++++++++++++++++++++++++++++++++++++++++++++++++++++++++++++++++++++++++
\begin{table}[t!]
\begin{center}
\begin{tabular}{ll@{\hspace{0.5cm}}l@{\hspace{0.5cm}}l@{\hspace{0.5cm}}l}
\hline
Temp.   & $L^3_s  \times L_t$ & $\beta$ & $a$  & $L_s a$ \\
 (MeV)  &                                 &              & (fm)  & (fm) \\
\hline
 265.9  & $54^3 \times 6$   &   5.890  &  0.1237  & 6.68 \\
 266.4  & $54^3 \times 6$   &   5.891  &  0.1235  & 6.67 \\
 266.9  & $54^3 \times 6$   &   5.892  &  0.1232  & 6.65  \\
 267.4  & $54^3 \times 6$   &   5.893   & 0.1230  & 6.64  \\
 268.0  & $54^3 \times 6$   &   5.8941 & 0.1227  & 6.63 \\
 268.5  & $54^3 \times 6$   &   5.895   & 0.1225  & 6.62  \\
 269.0  & $54^3 \times 6$   &   5.896   & 0.1223  & 6.60  \\
 269.5  & $54^3 \times 6$   &   5.897   & 0.1220  & 6.59 \\
 270.0  & $54^3 \times 6$   &   5.898   & 0.1218  & 6.58 \\
 271.0  & $54^3 \times 6$   &   5.900   & 0.1213  & 6.55 \\
 272.1  & $54^3 \times 6$   &   5.902   & 0.1209  & 6.53 \\
 273.1  & $54^3 \times 6$   &   5.904   & 0.1204   & 6.50 \\
            &                             &               &                & \\
 269.2  & $72^3 \times 8$   &   6.056    & 0.09163 & 6.60  \\
 270.1  & $72^3 \times 8$   &   6.058    & 0.09132 & 6.58  \\
 271.0  & $72^3 \times 8$   &   6.060    & 0.09101 & 6.55  \\
 271.5  & $72^3 \times 8$   &   6.061    & 0.09086 & 6.54 \\
 271.9  & $72^3 \times 8$   &   6.062    & 0.09071 & 6.53  \\
 272.4  & $72^3 \times 8$   &   6.063    & 0.09055 & 6.52  \\
 272.9  & $72^3 \times 8$   &   6.064    & 0.09040 & 6.51  \\
 273.3  & $72^3 \times 8$   &   6.065    & 0.09025 & 6.50  \\
 273.8  & $72^3 \times 8$   &   6.066    & 0.09010  & 6.49 \\
\hline
\end{tabular}
\end{center}
\caption{The lattice setup. The physical scale was defined from the string tension. 
              The values of $\beta$ were adjusted such that $L_s \, a \simeq 6.5 - 6.6$ fm.}
\label{tempsetup}
\end{table}
%+++++++++++++++++++++++++++++++++++++++++++++++++++++++++++++++++++++++++++
%+++++++++++++++++++++++++++++++++++++++++++++++++++++++++++++++++++++++++++
%+++++++++++++++++++++++++++++++++++++++++++++++++++++++++++++++++++++++++++

The simulations were performed on several asymmetric lattices $L^3_s \times L_t$ with a physical spatial volume $\sim (6.5$ fm$)^3$
and $L_t = 6$, $8$. We take the inverse of the lattice time extension $T = 1 / L_t$ in physical units as the definition for temperature.

In order to illustrate the behaviour of the various propagators at temperatures above $T_c$, we report firstly the results obtained
on $64^3\times6$ for $\beta=6.0$ ($T = 324$ MeV). Furthermore, we will also investigate the results of simulations using two sets of lattices close to the 
critical temperature (see Table~\ref{tempsetup}): (i) a set of coarser lattices with a lattice spacing $a \sim 0.12$ fm and $\beta \sim 5.9$;
(ii) a second set of finer lattices with lattice spacing $a \sim 0.09$ fm and $\beta \sim 6.0$.
Although the simulations with the coarser and finer lattices do not cover exactly the same range of temperatures,
they will allow us to estimate the effect due to the use of a finite lattice spacing.
Given the relatively large physical volumes (6.49 -- 6.68 fm)$^3$, we hope that the finite volume effects are small. 
Indeed, the studies of the gluon propagator at zero temperature~\cite{OliveiraSilvaOliveiraetal2014} suggests that, for sufficiently large volumes,
finite volume effects do not change significantly the propagator. 
In the present work, due to finite computing resources, no attempts are made to look at Gribov copies effects  -- see e.g.~\cite{Cucchieri1997,Silva2004,Sternbeck2013} and
references there in for results at zero temperature.

The gauge configurations were generated with the Chroma library \cite{Edwards05}. For the gauge fixing we use 
the Fourier accelerated steepest descent method described in \cite{Davies88}, which was implemented using Chroma 
and PFFT \cite{Pippig13} libraries. For each gauge configuration, the gauge fixing was stopped when $\theta \le 10^{-15}$. As
described in \cite{SilvaOliveiraetal2014}, $\theta$ stands for the lattice equivalent of the average value of $(\partial_\mu A^a_\mu)^2$ per site and color index.

For each of the ensembles reported in Tab.~\ref{tempsetup}, the gluon propagator was computed using 
100 gauge configurations. 
For the generation of the links a combined Monte Carlo sweep of 4 Cabbibo-Mariani heat bath and 7 overrelaxation sweeps was used; the 
measurements were performed each 100 combined sweeps after discarding the first 500 combined sweeps in the Markov chain 
for thermalisation.

\subsection{The gluon propagator}

The computation of the gluon propagator requires a definition of the $A_\mu$ from the links. In the current work we take the usual
expression
\be
   a \, g_0 \, A_\mu (x + \frac{a}{2} \hat{e}_\mu ) = \frac{1}{2 i}
    \left[\mathcal{U}_\mu (x) - \mathcal{U}^\dagger_\mu (x) \right]_{traceless} \ ,
   \label{Eq:gluon_field0}
\ee   
where $\hat{e}_\mu$ is the unit vector along the lattice direction $\mu$ and $g_0$ is the bare coupling constant.
The above definition assumes that the link and the gluon field are related by
\begin{eqnarray}
    \mathcal{U}_\mu (x) & = & \exp \left( i \, a \, g_0 \, A_\mu (x + \frac{a}{2} \hat{e}_\mu ) \right) \nonumber \\
    & & \approx      1 + i \, a \, g_0 \, A_\mu (x + \frac{a}{2} \hat{e}_\mu ) \ .
    \label{Eq_gluon_link_field0}
\end{eqnarray}
The above relations are certainly valid when one considers fluctuations around the trivial configuration, as in e.g. perturbation theory.
Given that we will look at configurations whose Polyakov loop is of type $|L|e^{i \theta}$ with $\theta \approx 0, \pm 2 \pi/3$ one might 
 ask whether
the above definition is still valid when $\theta = \pm 2 \pi/3$. 

\begin{figure*}[t]
\begin{center}
\includegraphics[angle=0,scale=0.45]{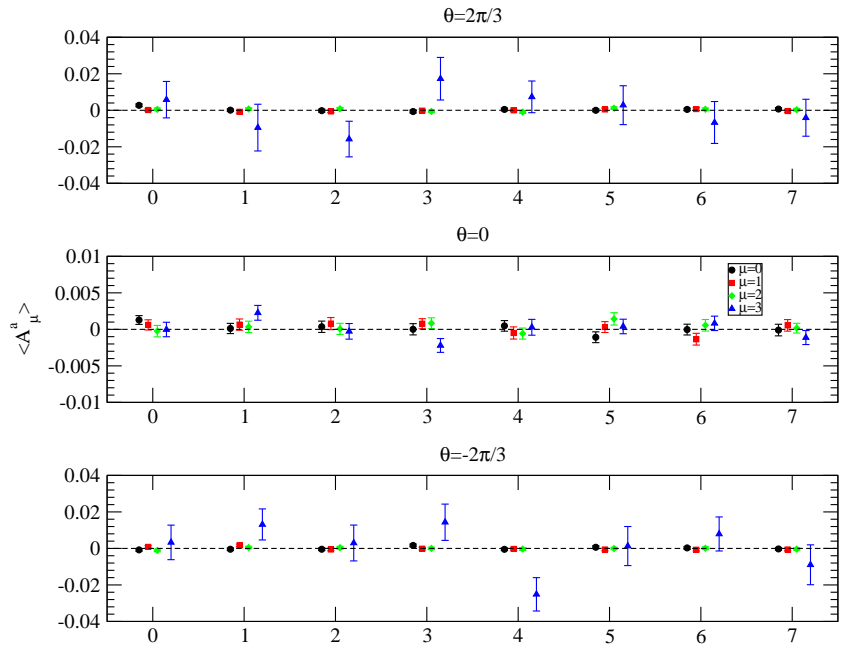}
\caption{Average values of $A^{a}_{\mu}(x)$ for a $64^3\times6$, $\beta=6.0$ ($T=324$ MeV) configuration.}
   \label{averageAx}
\end{center}
\end{figure*}

In Fig.~\ref{averageAx} we show the lattice average values of 
$A^a_{\mu}(x)$ for $a=0,\ldots,N_c^2-1$, $\mu=0\ldots N_d-1$ for a $64^3\times6$ configuration. In all cases,
$\langle A^a_\mu \rangle$  is compatible with zero within one or two standard deviations for all color and Lorentz indices. 
Furthermore, apart from a scaling factor, there is no clear difference between the configurations associated to the different phases of Polyakov loop.
We take this result as an indication that the gluon field given by  (\ref{Eq:gluon_field0}) can be applied to all configurations, including
all possible values of the Polyakov loop, considered here.

A generalized connection between the link and the gluon field  given by
\be
    \mathcal{U}_\mu (x) \approx u_0 \left[ 1 + i \, a\, g_0 \left(  A_\mu(x + \frac{a}{2} \hat{e}_\mu )  + a_\mu \right) \right] \ ,
    \label{Eq_gluon_field_general0}
\ee    
where $u_0$ is a real number and $a_\mu$ are constant fields, could be used. However, the replacement
of (\ref{Eq_gluon_link_field0}) by (\ref{Eq_gluon_field_general0})  gives the same bare gluon field up to a multiplicative
constant and a different zero momentum gluon field.
In what concerns  $u_0$, the use of a MOM scheme to renormalize the propagator removes any dependence on $u_0$.
It follows that the differences of using (\ref{Eq_gluon_link_field0}) or (\ref{Eq_gluon_field_general0}) to compute the gluon propagator
can only appear for the zero momentum propagator, which leaves unchanged the main conclusions of the current work.

In the Landau gauge and at finite temperature, the gluon propagator reads
\be
\langle A_\mu^a(p) A_\nu^b(q)\rangle = V \, \delta^{ab} \, \delta( p+q) \, D^{ab}_{\mu\nu}(p) \ ,
\ee
where 
\be
D^{ab}_{\mu\nu}(p)  =  \delta^{ab}\Bigg\{ P^{T}_{\mu\nu} \, D_{T}(p_4,\vec{p}) + P^{L}_{\mu\nu} \, D_{L}(p_4,\vec{p}) \Bigg\} \ ,
\label{tens-struct}
\ee
and the  transverse and longitudinal projectors are given by
\bea
P^{T}_{\mu\nu} &=& (1-\delta_{\mu 4})(1-\delta_{\nu 4})\left(\delta_{\mu \nu}-\frac{p_\mu p_\nu}{\vec{p}^{\,\, 2}}\right) \quad , 
\\ %\non
P^{L}_{\mu\nu} &=& \left(\delta_{\mu \nu}-\frac{p_\mu p_\nu}{{p}^{2}}\right) - P^{T}_{\mu\nu} \, .
\label{long-proj}
\eea
In the above expressions, Latin letters stand for color indices and Greek letters for space-time indices. 

The results shown  here are for renormalized longitudinal and transverse propagators. For the
renormalization we follow the procedure devised in~\cite{SilvaOliveiraetal2014}, taking $\mu = 4$ GeV for the renormalization scale
and setting $D_{L,T}(\mu^2) = Z_R D^{Lat}_{L,T} (\mu^2) = 1/\mu^2$. The longitudinal and transverse form factors
were renormalized independently within each possible phase value of the associated Polyakov loop. 
It turns out that, in our simulations, the renormalization constants for the longitudinal and transverse form factors agree within one standard deviation 
for all possible values of the phase.

\subsection{Center Symmetry}

For pure Yang-Mills theory formulated on the lattice, the Wilson action and the path integral measure are invariant under the SU(3)
group center, i.e. under transformations of type
\be
   \mathcal{U}_4 ( \vec{x}, t = 0) \longrightarrow 
   \mathcal{U}^\prime_4 ( \vec{x}, t = 0) = z \,\mathcal{U}_4 ( \vec{x}, t = 0)
   \label{Eq:center_trans0} 
\ee
for all $z \in Z_3$. For temperatures below $T_c$ the center symmetry is preserved and the various phases of
$L$ are sampled likewise which implies  $\langle L \rangle \sim 0$. Above the critical temperature, the center symmetry 
is spontaneously broken and the average value of $L$ over the lattice no longer vanishes. 

As described in \cite{Gattringer10,Gattringer11,Endrodi14}, for $T > T_c$ it is possible to identify center domains where the 
phase of the Polyakov loop is $ \approx 0, \, \pm \,2 \, \pi /3$, i.e. it coincides with the phase of the elements of $Z_3$.
The dimensions of the center domains are
temperature dependent and, at the critical temperature, these clusters percolate the lattice. In this sense, the gauge
configurations can be classified according to the associated phase of the Polyakov loop,
\begin{equation}
    L = \langle L \rangle = |L| \, e^{i \theta} \ .
\end{equation}
Center transformations map configurations in different equivalent classes, i.e. with different $\theta$. 
We have performed a  number of simulations using various volumes, results not shown here, and they suggest that, in the Markov chain,
the probability for the transition between the equivalent classes decreases
when the physical volume of the spatial lattice increases. This suggests that, in the limit of infinite volume, the sampling is confined to configurations whose Polyakov loop
is such that  $\theta$ takes one and only one value in $\{0, \pm 2 \pi/3\}$.

\subsection{$Z_3$ Sectors}

\begin{figure}[t]
\centering
\begin{tikzpicture}[>=latex]

\foreach \ang/\lab/\dir in {
%  0/0/right,
  120/{2 \pi/3}/{above right},
  240/{-2\pi/3}/below} {
  \draw[ultra thick] (0,0) -- (\ang : 2.5);
  \node [fill=white] at (\ang : 2.5) [\dir] {\scriptsize $\lab$};
}

\draw[ultra thick] (0,0) -- (0 : 2.5);

\draw[->] (0,0) -- (90 : 3);
\node at (0.5,2.5) {Im $L$};

\draw (0,0) -- (180 :3);
\draw[->] (0,0) --  (0 : 3);
\node at (2.5,-0.3) {Re $L$};

\shadedraw[inner color=yellow,outer color=orange,draw=white] (0,0) -- (0.75,-1.30) arc (-60:60:1.5) -- cycle;
\shadedraw[inner color=white,outer color=orange,draw=white] (0,0) -- (0.75,1.30) arc (60:180:1.5) -- cycle;
\shadedraw[inner color=orange,outer color=yellow,draw=white] (0,0) -- (0.75,-1.30) arc (-60:-180:1.5) -- cycle;

\draw[ultra thick,dotted] (0,0) -- (60:2);
\draw[ultra thick,dotted] (0,0) -- (-60:2);
\draw[ultra thick,dotted,<->] (0.75,-1.30) arc (-60:60:1.5);
\draw[ultra thick,dotted,<->] (0.75,1.30) arc (60:180:1.5);
\draw[ultra thick,dotted,<->] (0.75,-1.30) arc (-60:-180:1.5);

\node [fill=white] at (1.8,1.3) {Sector $0$};
\node [fill=white] at (-2.,-1.3) {Sector $-1$};
\node [fill=white!50] at (-2.,1.3) {Sector $1$};

%\fill[green!20!white] (0,0) -- (0.75,-1.30) arc (-60:60:1.5) -- cycle;
%\shadedraw[inner color=yellow,outer color=black,draw=yellow] (6,0) rectangle +(2,1);
\end{tikzpicture} 
\caption{The definition of the $Z_3$ sectors.}
\label{Fig:Polyakov_sectors0}
\end{figure}
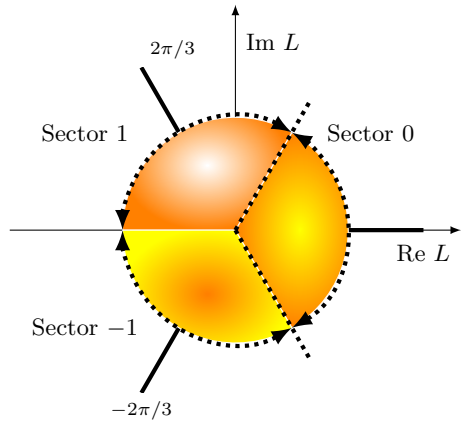

Our goal is to try to understand if the dynamics of the gluon field changes with the phase of the Polyakov loop. 
The transformations of type (\ref{Eq:center_trans0}) map configurations with the same action  which, from the 
point of view of the sampling, 
belong to classes with exactly the same probability. In principle, one
could include these transformations in the definition of the Markov process and, in this way, sample equally all the
possible phases of the Polyakov loop. However,  in this work the 
gauge configurations are generated in the usual way
and, before 
gauge fixing, to each configuration one applies transformations of type (\ref{Eq:center_trans0}) for all $z \in Z_3$.
Then, each of these configurations is rotated to the (minimal) Landau gauge and classified according to the phase of the 
corresponding average value of the Polyakov loop $ \langle L \rangle = |L| e^{i \theta} $ as
\be
    \theta = \left\{ \begin{array}{cll}
                           - \pi < \theta \le -\frac{\pi}{3} ,  & \hspace{0.5cm} & \mbox{Sector -1} , \\
                           & & \\
                            -\frac{\pi}{3}  < \theta \le \frac{\pi}{3} , &  & \mbox{Sector 0} , \\
                            & & \\
                            \frac{\pi}{3} < \theta \le \pi , &  & \mbox{Sector 1} \, .
                           \end{array} \right.
\ee    
This classification of the configuration is resumed in Fig.~\ref{Fig:Polyakov_sectors0}. We recall that, for each gauge configuration, although the Polyakov loop 
main contribution comes from  center domains belonging to a given $Z_3$ sector, the other sectors are also present
and belong to smaller center domains.

\begin{figure}[t]
\begin{center}
\begin{subfigure}{\columnwidth}
   \includegraphics[angle=-90,scale=0.24]{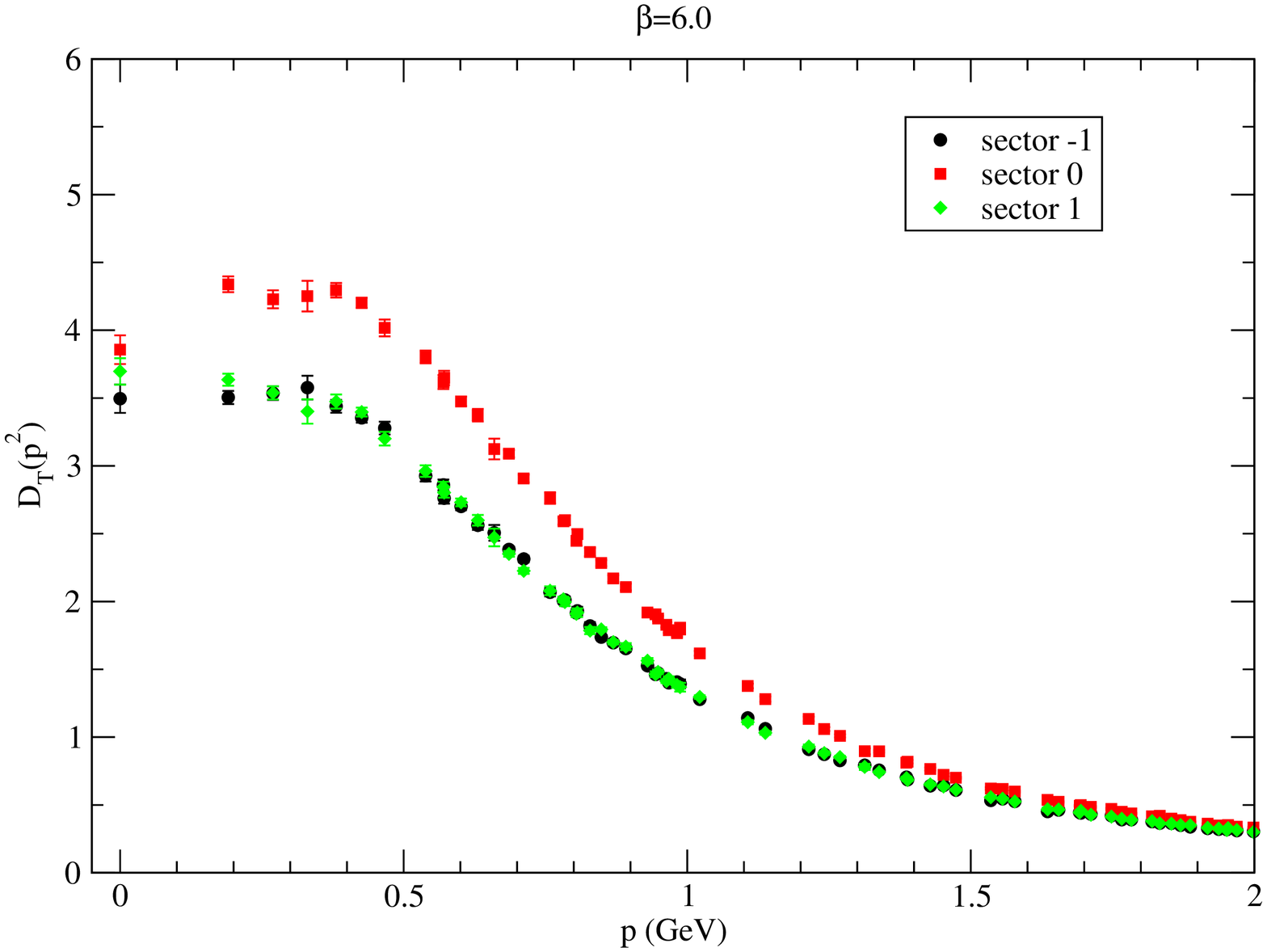}
     \caption{Magnetic Form Factor} 
    \end{subfigure}  
\begin{subfigure}{\columnwidth}
  \includegraphics[angle=-90,scale=0.24]{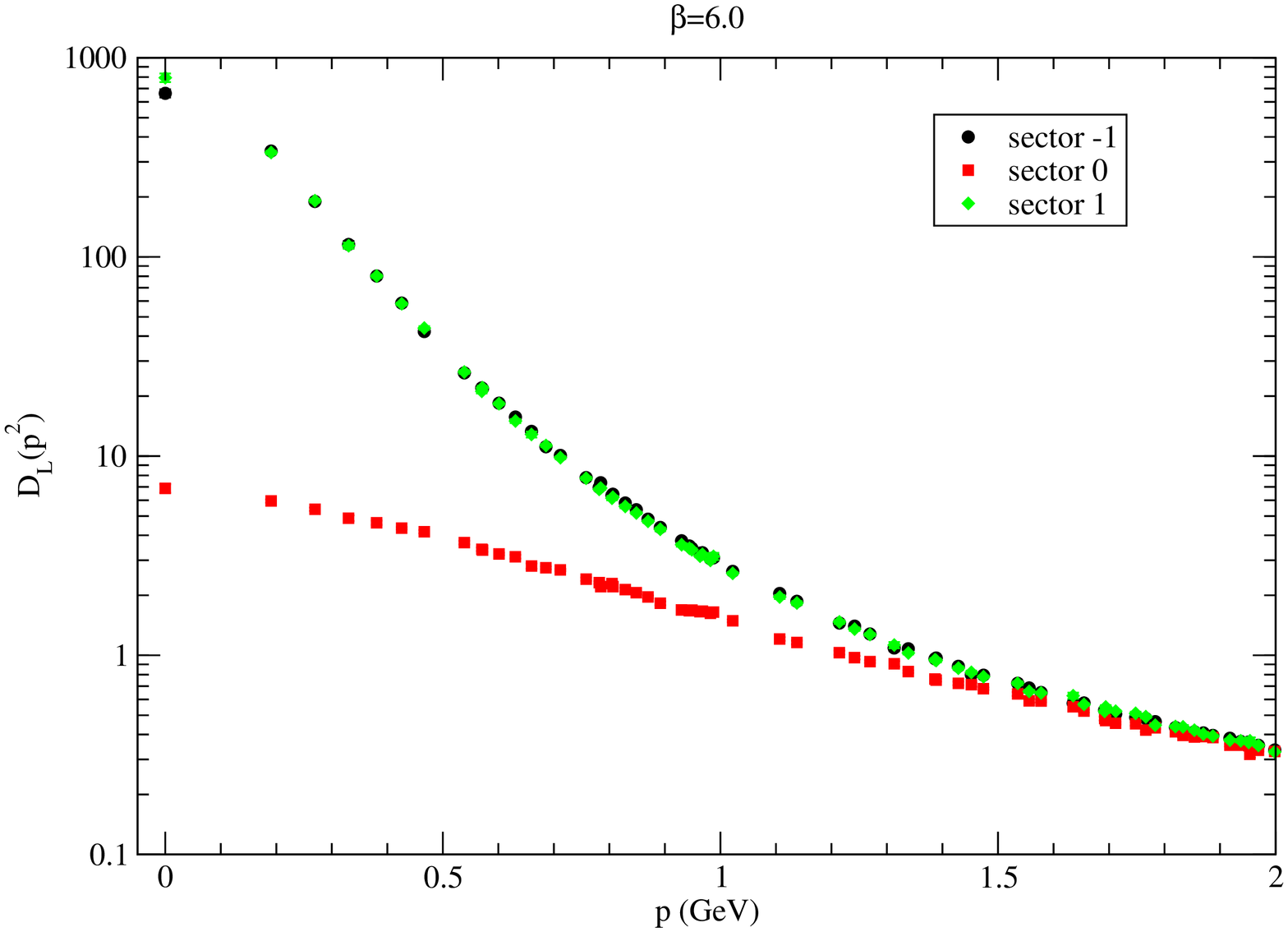} 
     \caption{Electric Form Factor} 
    \end{subfigure} 
\end{center}
  \caption{Gluon propagators for the different sectors at $T = 324$ MeV.}
   \label{temp324}
\end{figure}

In order to illustrate what happens to the gluon propagator in each of the $Z_3$ sectors, in Fig.~\ref{temp324} one shows the electric and magnetic components 
of the propagator per $Z_3$ sector for $T = 324$ MeV computed on a $64^3\times6$ lattice for $\beta=6.0$. 
Fig.~\ref{temp324} shows the typical behaviour of the propagators  for $T>T_c$.
For temperatures close to $T_c$ a
careful analysis is required, see discussion below, but the pattern observed in Fig.~\ref{temp324} still applies if we approach the critical temperature from above.

If the electric form factor for the 0 sector is suppressed relative to
the $\pm 1$ sectors, for the magnetic form factor the situation is reversed with the sector 0 being enhanced relative to the $\pm 1$ sectors. 
One can translate this result into a mass scale defined by the inverse of the square root of the propagator  at zero momentum. The mass scale
associated with the electric sector is much larger for the 0 sector, in comparison  with the $\pm 1$ sectors. On the other hand,
the mass associated with the magnetic sector is smaller for the 0 sector, in comparison  with the $\pm 1$ sectors. From Fig.~\ref{temp324}
one can identify the following mass
hierarchy $m_L ( \pm 1) < m_L ( 0) < m_M (0) < m_M ( \pm 1 )$, where $m_L$ ($m_M$) stands for electric (magnetic) mass
and, in parenthesis, one identifies the corresponding sector.

%=============================================================================
%=============================================================================
\section{Simulations near the phase transition \label{sec3}}

\begin{figure*}[t]
   \begin{center}
      \begin{subfigure}{\columnwidth}
       \includegraphics[scale=0.20,angle=-90]{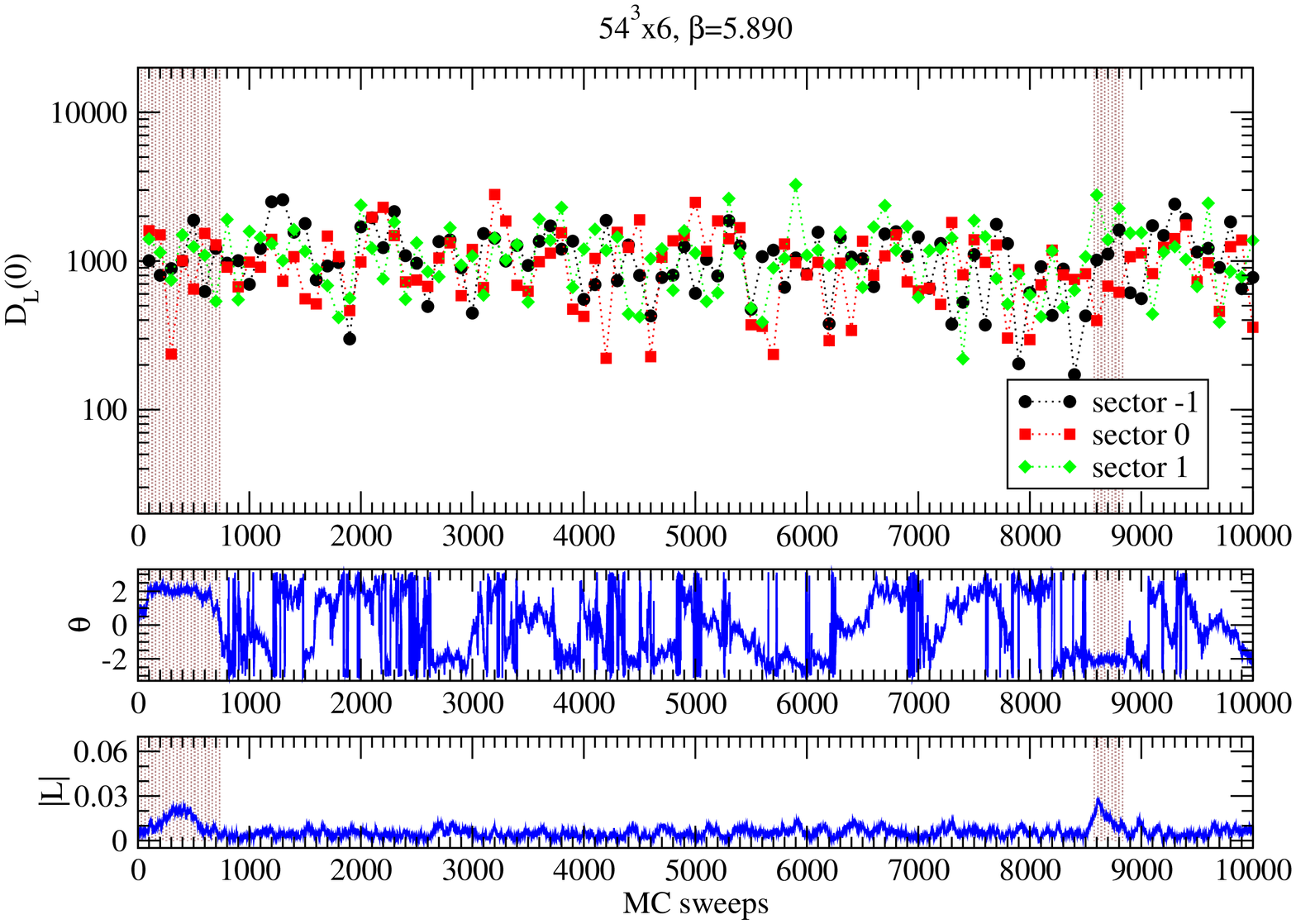} 
                        \caption{$T = 265.9$ MeV }
      \end{subfigure} \hfill
      \begin{subfigure}{\columnwidth}
      \includegraphics[scale=0.20,angle=-90]{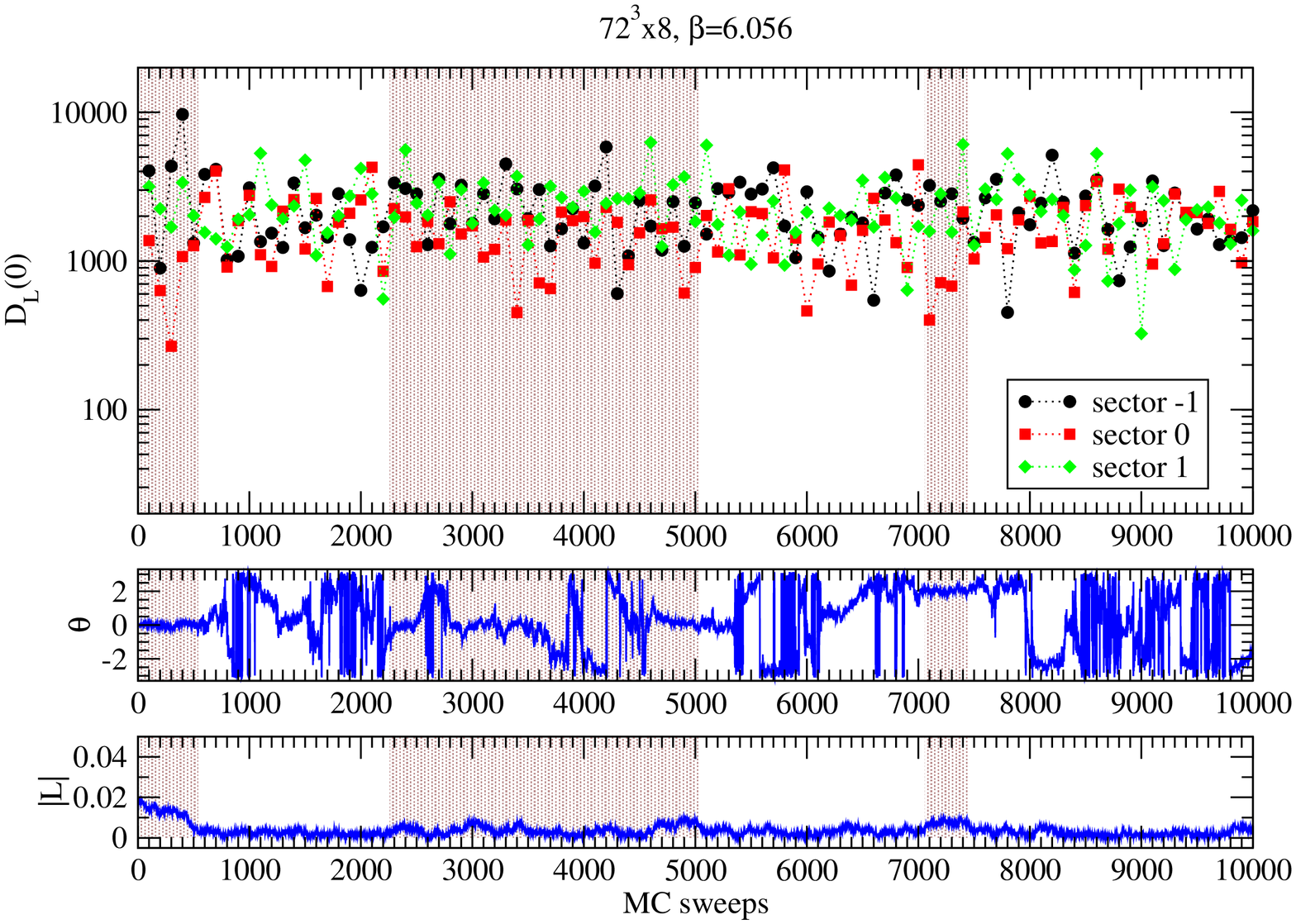} 
                       \caption{$T = 269.2$ MeV} 
                       \end{subfigure} \\
%     \vspace{0.7cm}
     \begin{subfigure}{\columnwidth}
      \includegraphics[scale=0.20,angle=-90]{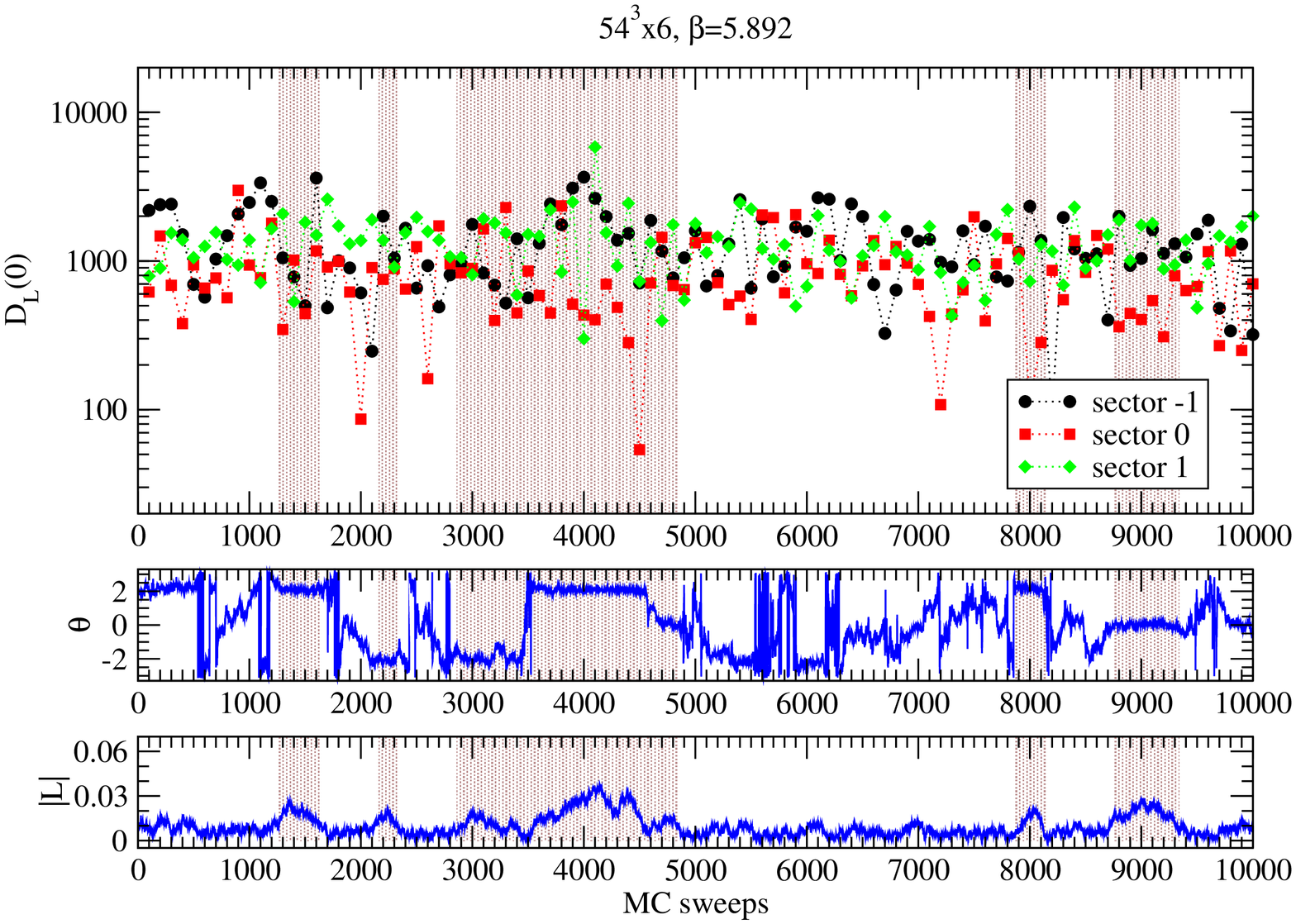} 
                       \caption{$T = 266.9$ MeV}
                       \end{subfigure} \hfill
     \begin{subfigure}{\columnwidth}
      \includegraphics[scale=0.20,angle=-90]{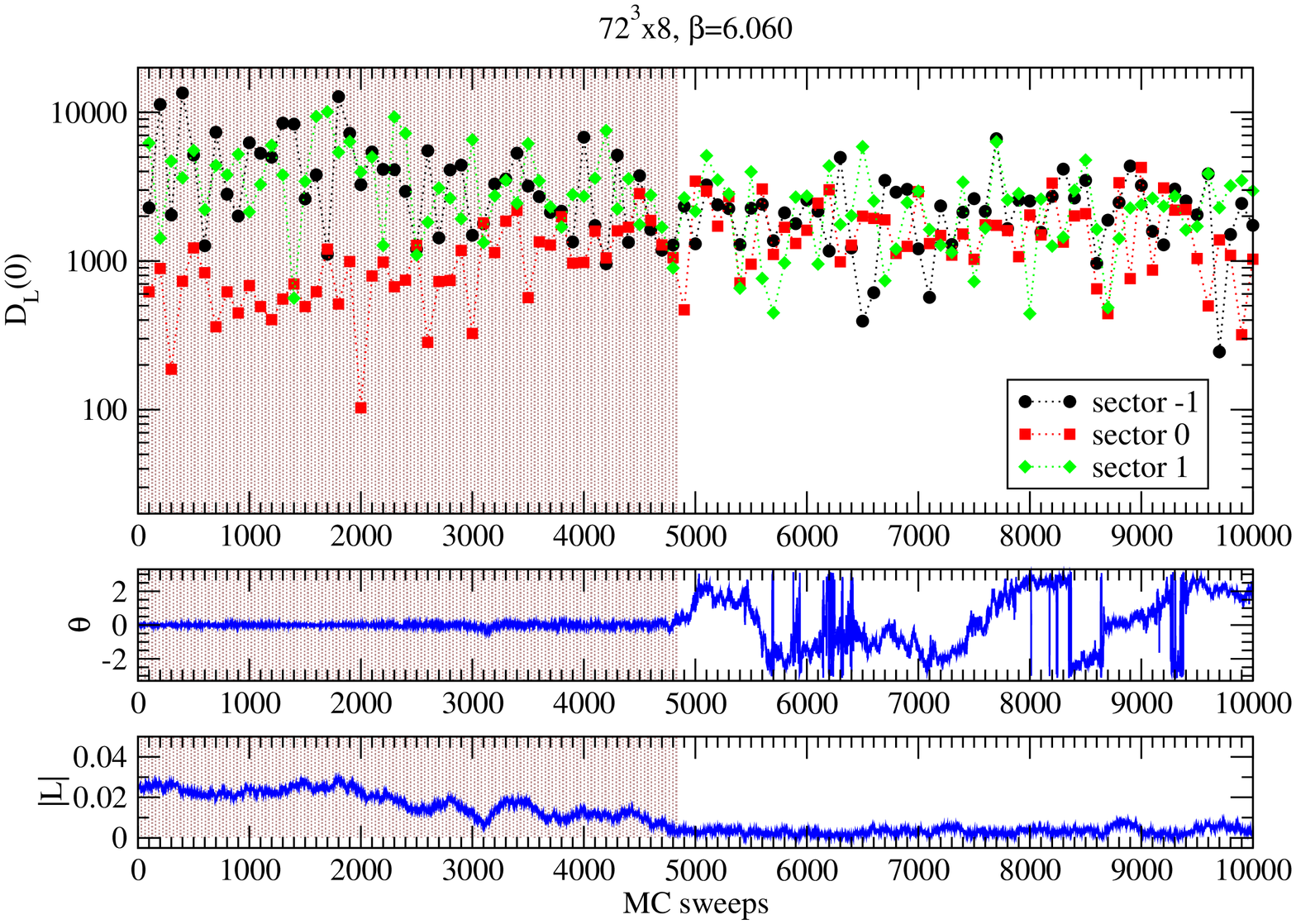} 
                       \caption{$T = 271$ MeV} 
                       \end{subfigure} \\
%     \vspace{0.7cm}
     \begin{subfigure}{\columnwidth}
      \includegraphics[scale=0.20,angle=-90]{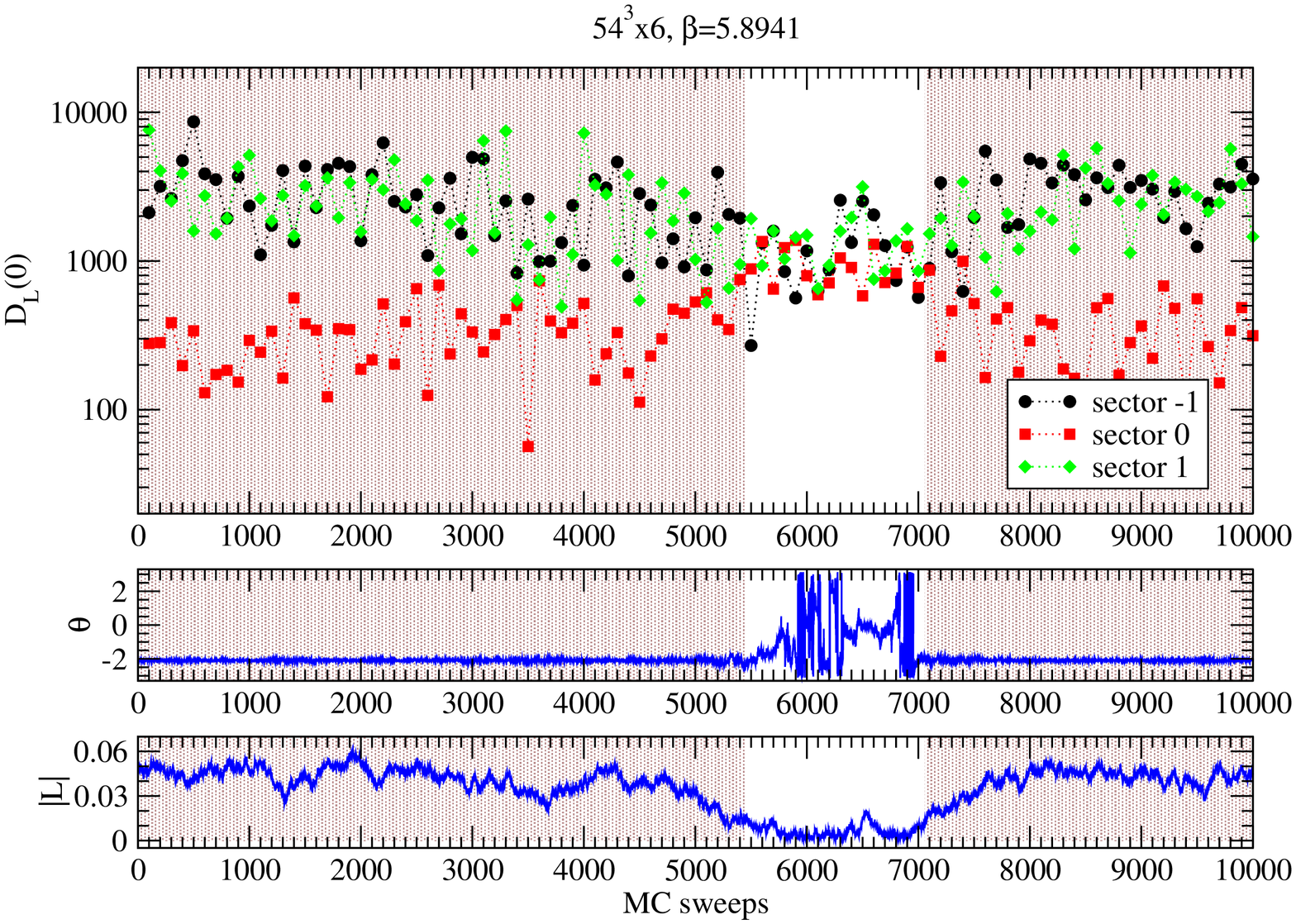} 
                       \caption{$T = 268$ MeV}
                       \end{subfigure} \hfill
     \begin{subfigure}{\columnwidth}
      \includegraphics[scale=0.20,angle=-90]{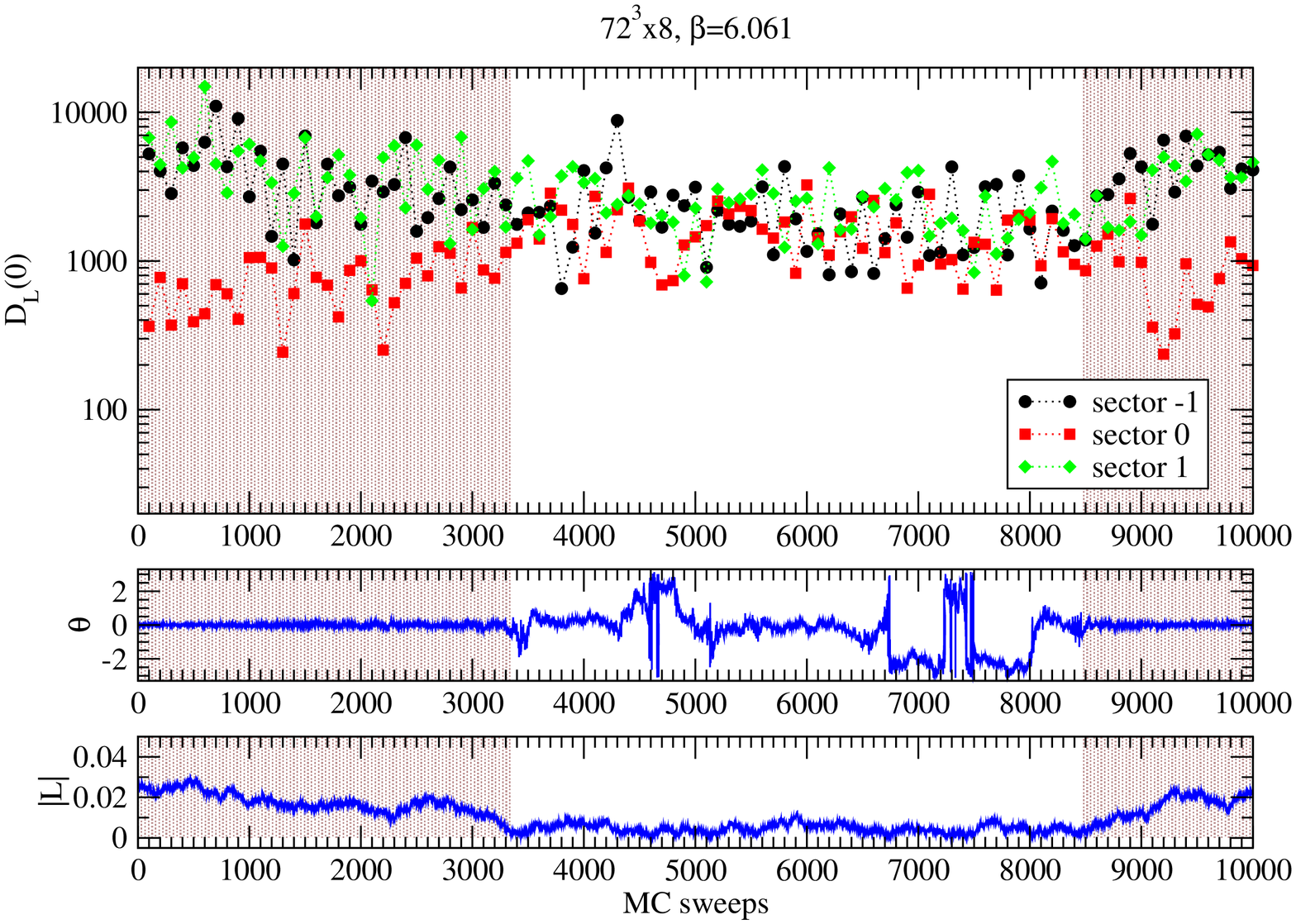} 
                       \caption{$T = 271.5$ MeV} 
                       \end{subfigure} 
%     \vspace{0.7cm}
     \begin{subfigure}{\columnwidth}
      \includegraphics[scale=0.20,angle=-90]{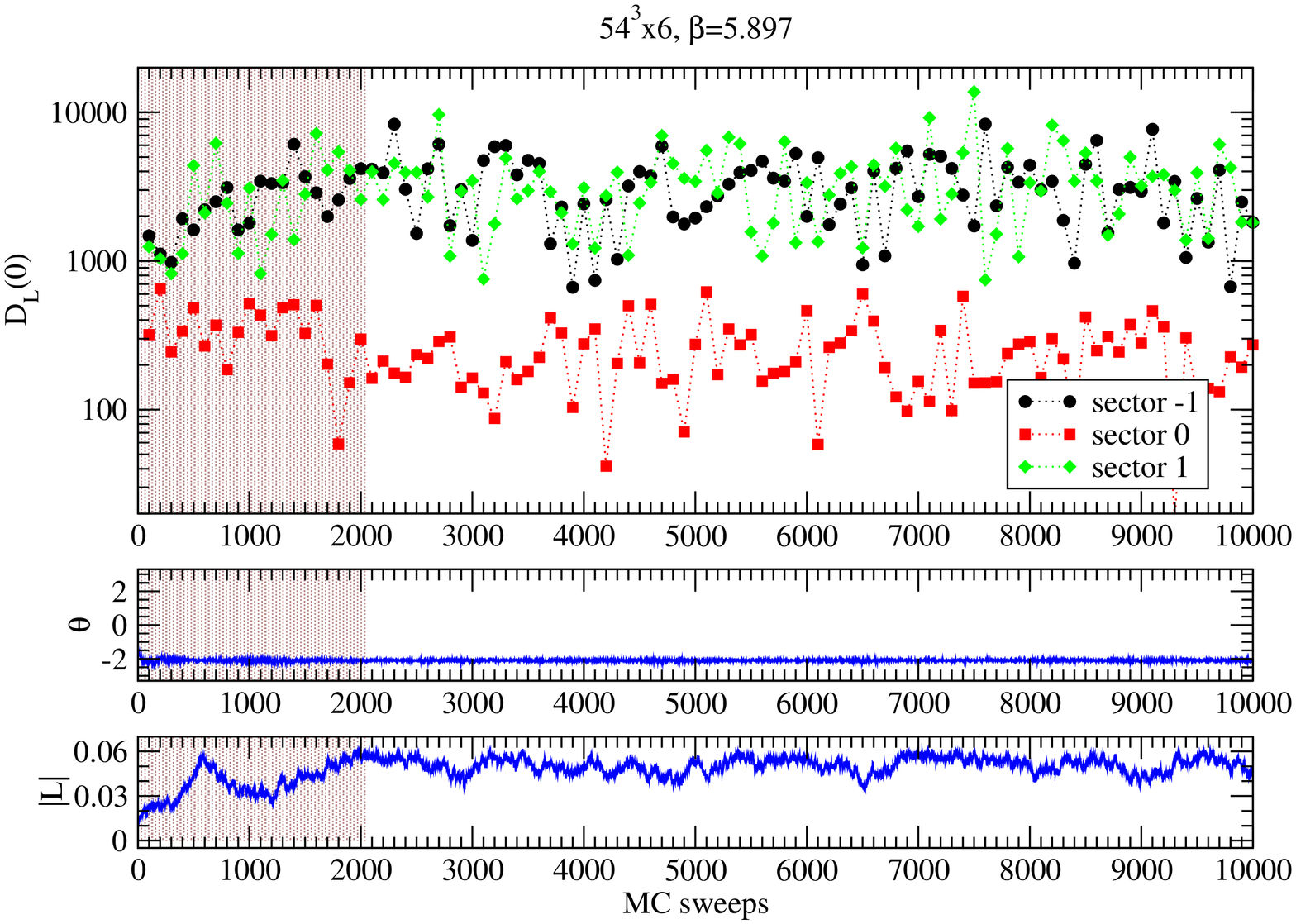} 
                       \caption{$T = 269.5$ MeV}
                       \end{subfigure} \hfill
     \begin{subfigure}{\columnwidth}
      \includegraphics[scale=0.20,angle=-90]{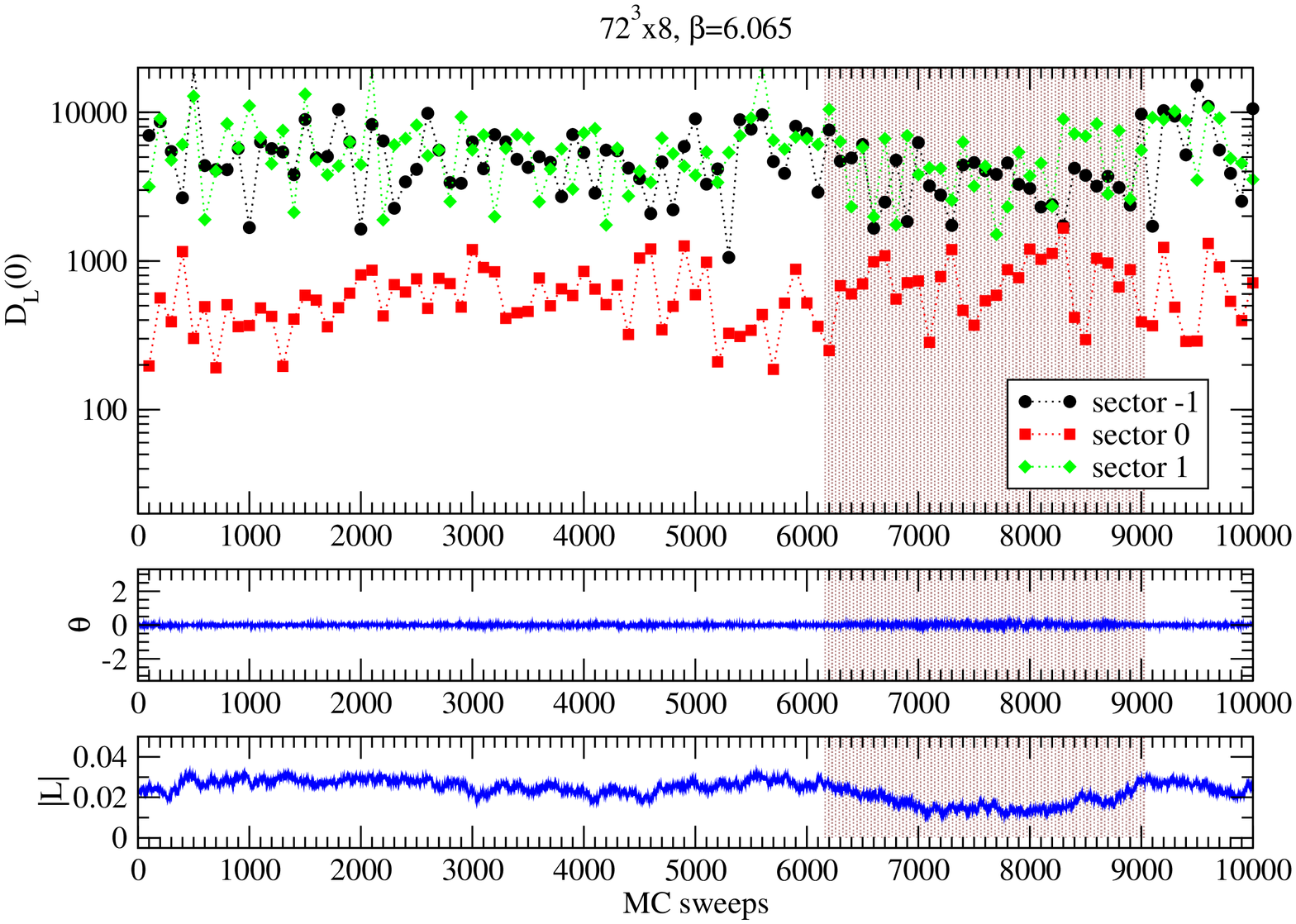} 
                       \caption{$T = 273.3$ MeV} 
                       \end{subfigure} 
  \end{center}
   \caption{Monte Carlo history of the coarse (left) and fine (right) ensembles for bare $|L|$, $\theta$ and $D_L(0)$. Configurations inside the colored areas will not be used to compute the propagators; see text for details. }
   \label{fig:D0L_T_all}
\end{figure*}

The differences observed in the gluon propagator form factors and reported in Fig.~\ref{temp324} seems to be a feature of the deconfined phase 
and happens for $T > T_c$. 
For the pure glue theory, the deconfinement phase transition is of first order. 
Near the critical temperature, Monte Carlo simulations access configurations in the confined and deconfined phase. 
The simulations  we have performed for $T \approx T_c \approx 270$ MeV show also
that the probability transition between the confined and deconfined phase in the Markov chain decreases  when we increase the physical volume used to simulate the theory. 
In order to properly describe any  of the two phases near $T_c$, one needs to define a way to separate the configurations belonging to either phases.

In Fig.~\ref{fig:D0L_T_all} we report the Markov chain history for simulations performed with a temperature close to $T_c$ and for the coarser and
finer lattices. The figure includes the corresponding values of the modulus of the bare Polyakov loop, its phase and the longitudinal (electric) gluon form factor.
For the simulations reported, the modulus of the bare Polyakov loop seems to take a continuous range of values. 
When it takes higher values, one observes
that $D_L(0)$ computed for configurations in $\pm 1$ $Z_3$ sectors differs substantially from the propagator in the zero sector. On the other hand,
when $|L|$ takes smaller values, it follows that the propagators computed using configurations in any of the $Z_3$ sectors are indistinguishable.  We associated the first
type of configurations with the deconfined phase, while the latest family of configurations has been identified
with the confined phase. Recall that in the results reported in Fig.~\ref{temp324}, for temperatures well above the critical temperature, it was observed
that $D_L(0)$ for $Z_3$ sectors $\pm 1$ is enhanced relative to the zero sector. Furthermore, the time evolution of $\theta$ shows also that for
lower values of $|L|$, the phase of the Polyakov loop fluctuates freely, suggesting a $|L| \sim 0$ as expected in the confined phase.

The observed correlations in the time evolution of $|L|$, $\theta$ and the differences in $D_L(0)$ associated to the various $Z_3$ sectors, suggests
that this difference between the propagators can be used as a criterion to identify the phase, confined or deconfined, of a given
configuration. 
In this spirit,
the simulations performed on the coarser lattices with $\beta\leq5.895$ ($T \leq 268.5$ MeV), or on finer lattices and $\beta\leq6.061$ ($T \leq 271.5$ MeV) provide, in general, configurations in the confined phase.
On the other hand, simulations with higher values of $\beta$ are mainly in the deconfined phase.

In Fig.~\ref{fig:D0L_T_all} we try to separate the various configurations using the criterion discussed above.  For simulations on coarse lattices with $\beta\leq5.895$, and finer lattices with $\beta\leq6.061$, the configurations identified with the confined phase are plotted in the figure against a white background, with all the others, including those which are not clearly in any of the phases, are plotted against a colored background. For the other simulations with higher $\beta$ and $T$, the white background identifies configurations in the deconfined phase, while the colored background refers to all the others.
The configurations associated to the dominant phase in a given simulation are plotted against a white
background; the exception to this rule being the simulation for the coarser lattice with $\beta = 5.8941$ ($T = 268$ MeV). 
In fact, we found that the configurations generated in the simulation with $\beta=5.895$ ($T = 268.5$ MeV) are mainly in the confined phase, while the configurations 
generated with $\beta=5.896$ ($T = 269$ MeV) are all in the deconfined phase. On the other hand, for the simulation using the intermediate value $\beta=5.8941$ 
($T = 268$ MeV)
most of the configurations are in the deconfined phase and, therefore, for this intermediate $\beta$ we will use in our calculations only the configurations in the confined phase.

The use of the separation of $D_L(0)$ to identify confined and deconfined configurations in the Monte Carlo time history allows an estimate of $T_c$.
The simulations point towards a $T_c$ in the range $269-272$ MeV, in good accordance with the literature.

\begin{figure*}[t]
   \begin{center}
      \begin{subfigure}{\columnwidth}
      \includegraphics[scale=0.22,angle=-90]{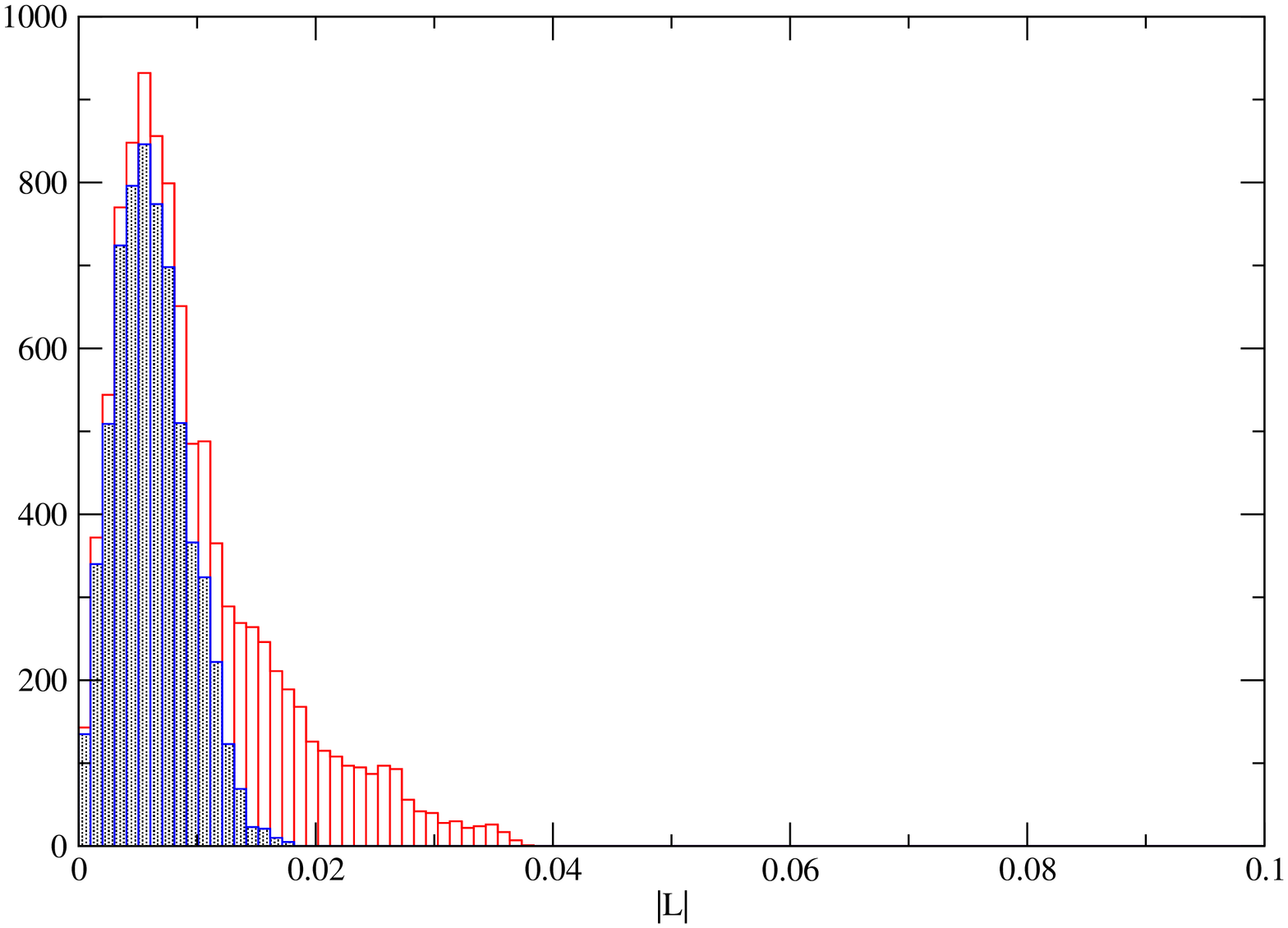} 
         \caption{coarser lattice for $T = 266.9$ MeV}   
      \end{subfigure} 
     \hfill
     \begin{subfigure}{\columnwidth}
      \includegraphics[scale=0.22,angle=-90]{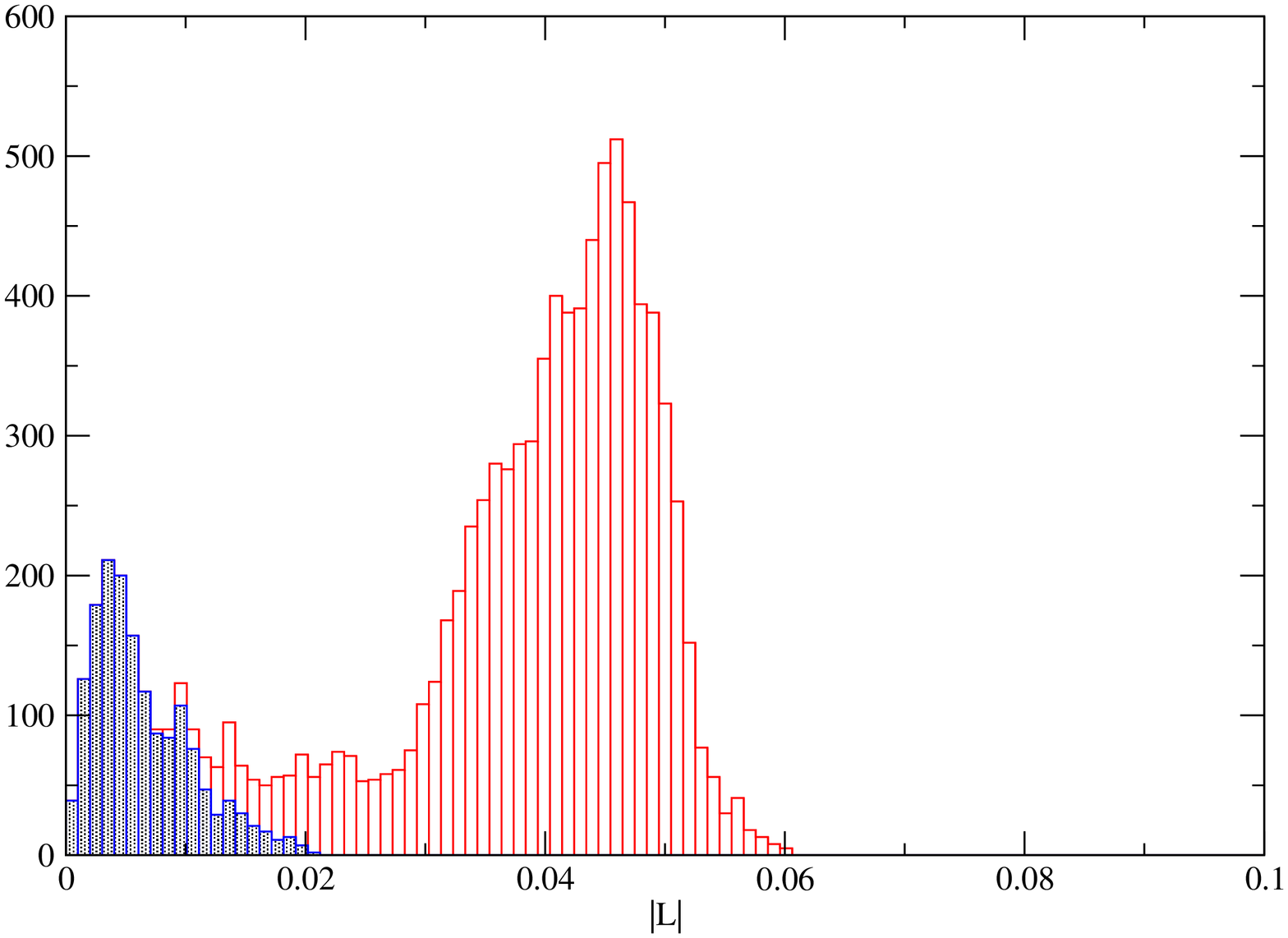} 
        \caption{coarser lattice for $T = 268$ MeV}
       \end{subfigure} \\
     \begin{subfigure}{\columnwidth}
      \includegraphics[scale=0.22,angle=-90]{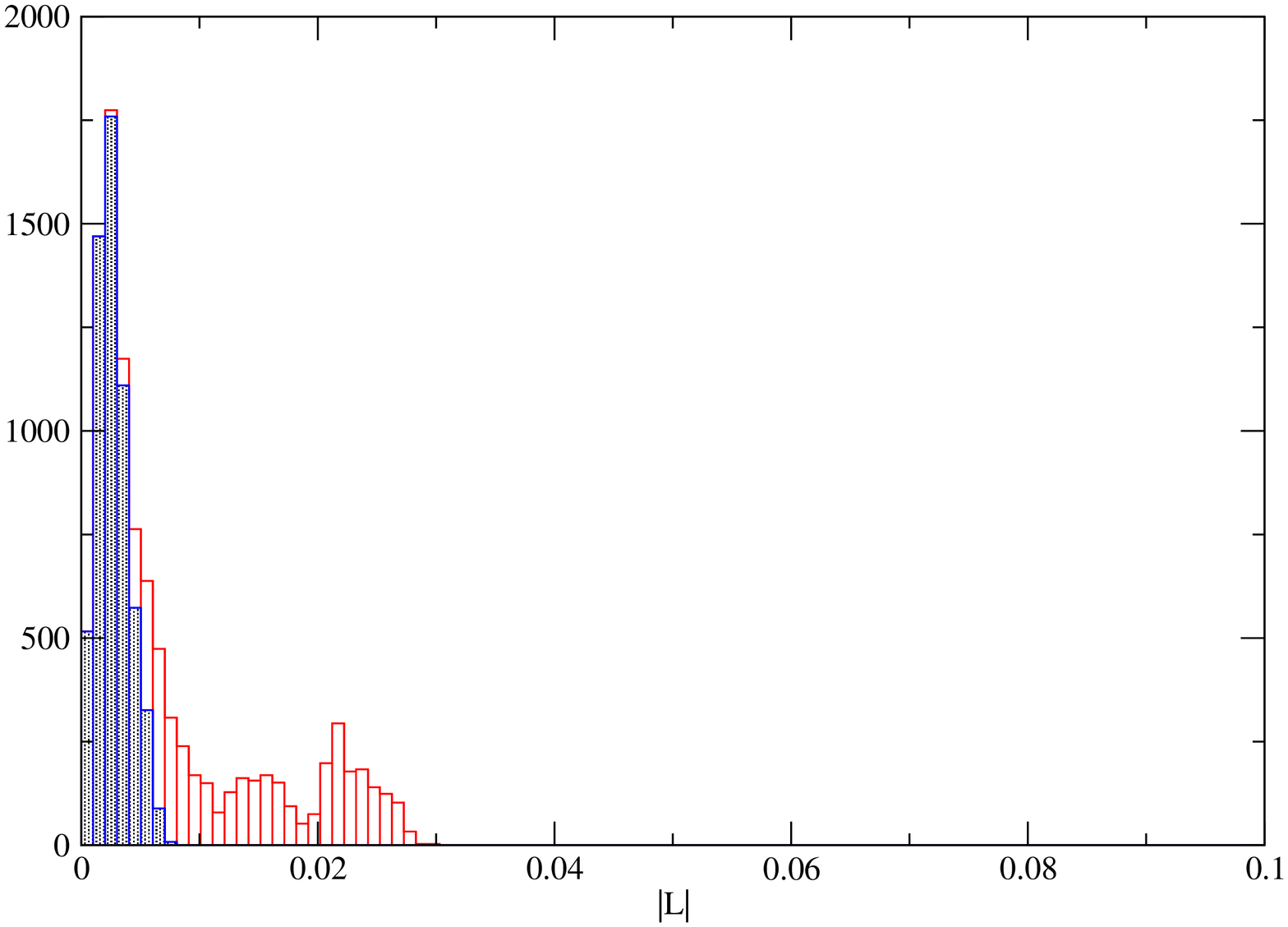} 
         \caption{Finer lattice for $T = 270.1$ MeV}
       \end{subfigure} \hfill
     \begin{subfigure}{\columnwidth}
      \includegraphics[scale=0.22,angle=-90]{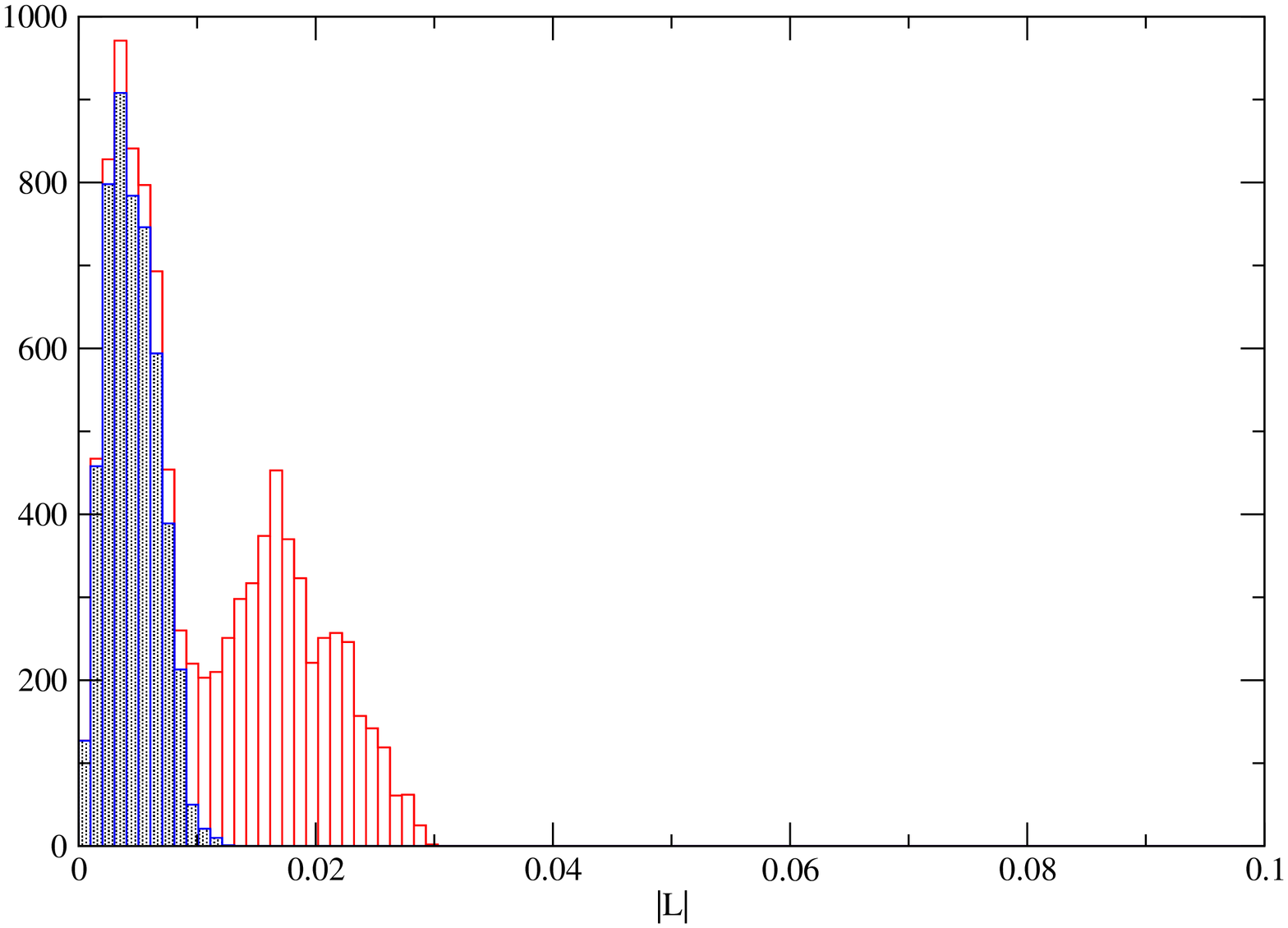} 
      \caption{Finer lattice for $T = 271.5$ MeV}
     \end{subfigure} 
  \end{center}
   \caption{Histograms of $|L|$ for various $T$ before (red) and after the selection procedure (blue) for all independent configurations obtained during the sampling. See text for details.}
   \label{fig:polyclean}
\end{figure*}

\begin{figure*}[t]
   \begin{center}
     \begin{subfigure}{\columnwidth}
      \includegraphics[scale=0.26,angle=-90]{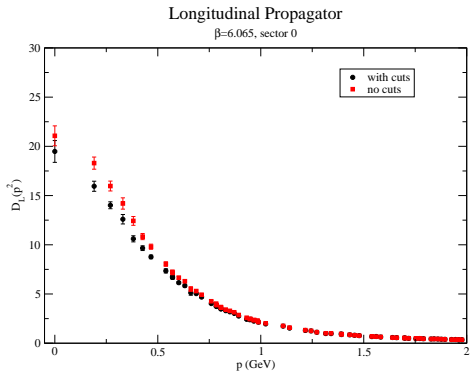} 
      \caption{$L=72$, $\beta=6.065$, sector 0.}
     \end{subfigure} \hfill
     \begin{subfigure}{\columnwidth}
      \includegraphics[scale=0.26,angle=-90]{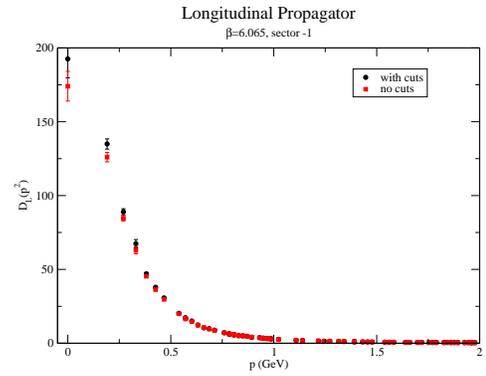} 
      \caption{$L=72$, $\beta=6.065$, sector -1.} 
     \end{subfigure} 
  \end{center}
   \caption{Example of how the selection procedure change the gluon electrical form factor for the finer lattice and $T=273.3$ MeV.}
   \label{fig:cutprop}
\end{figure*}

In order to illustrate the effects of the selection procedure just described, in Fig~\ref{fig:polyclean} we show the histograms of the modulus of the Polyakov loop before (red lines) and after (blue lines) removing 
the configurations that, according to our criterion, will not be used to compute the propagators. If one considers all the configurations generated by the sampling, the
histograms spread over much larger values of $|L|$ and often show several maxima. On the other hand, our selection of configurations gives rise to
a histogram where the distribution of the values of the Polyakov loop clearly have a single maximum. To illustrate the effects on the gluon propagator,
in Fig.~\ref{fig:cutprop} we report an example where the longitudinal gluon form factor is shown, before and after performing our selection of configurations. As can be seen, the effect of our selection can be well beyond one standard deviation.

%================================================================
%================================================================
\section{The Gluon Propagator Near the Phase Transition \label{sec4}}

%+++++++++++++++++++++++++++++++++++++++++++++++++++++++++++++++++++++++++++
%+++++++++++++++++++++++++++++++++++++++++++++++++++++++++++++++++++++++++++
%+++++++++++++++++++++++++++++++++++++++++++++++++++++++++++++++++++++++++++
\begin{table}[t!]
\begin{center}
\begin{tabular}{ll@{\hspace{0.5cm}}l@{\hspace{0.5cm}}l}
\hline
Temp.   & $L^3_s  \times L_t$ & $\beta$ & $\# Configs$ \\
 (MeV)  &                                 &              &  \\
\hline
 265.9  & $54^3 \times 6$   &   5.890  &  90  \\
 266.4  & $54^3 \times 6$   &   5.891  &  91  \\
 266.9  & $54^3 \times 6$   &   5.892  &  65   \\
 267.4  & $54^3 \times 6$   &   5.893   & 76   \\
 268.0  & $54^3 \times 6$   &   5.8941 &  16   \\
 268.5  & $54^3 \times 6$   &   5.895   & 63    \\
 269.0  & $54^3 \times 6$   &   5.896   & 100    \\
 269.5  & $54^3 \times 6$   &   5.897   & 80  \\
 270.0  & $54^3 \times 6$   &   5.898   & 100   \\
 271.0  & $54^3 \times 6$   &   5.900   & 95  \\
 272.1  & $54^3 \times 6$   &   5.902   & 100   \\
 273.1  & $54^3 \times 6$   &   5.904   & 100    \\
            &              &            & \\
 269.2  & $72^3 \times 8$   &   6.056    & 63   \\
 270.1  & $72^3 \times 8$   &   6.058    & 59   \\
 271.0  & $72^3 \times 8$   &   6.060    & 52   \\
 271.5  & $72^3 \times 8$   &   6.061    & 51  \\
 271.9  & $72^3 \times 8$   &   6.062    & 82   \\
 272.4  & $72^3 \times 8$   &   6.063    & 70   \\
 272.9  & $72^3 \times 8$   &   6.064    & 100   \\
 273.3  & $72^3 \times 8$   &   6.065    & 71   \\
 273.8  & $72^3 \times 8$   &   6.066    & 100   \\
\hline
\end{tabular}
\end{center}
\caption{The same as Tab.~\ref{tempsetup}. The number of configurations refers to the configurations which, according to our selection procedure, will be used to compute the propagator. See text for details.}
\label{tempsetupclean}
\end{table}
%+++++++++++++++++++++++++++++++++++++++++++++++++++++++++++++++++++++++++++
%+++++++++++++++++++++++++++++++++++++++++++++++++++++++++++++++++++++++++++
%+++++++++++++++++++++++++++++++++++++++++++++++++++++++++++++++++++++++++++

As discussed in the last section, the proper computation of the gluon propagator for temperatures near $T_c$ needs a selection procedure to separate those configurations which can be associated to the 
deconfined or confined phase. This has been done looking at the Monte Carlo time evolution for the 
Polyakov loop and/or $D_L(0)$. Recall that we considered 100 configurations for each simulation, and our selection procedure implies that in several cases we will not use the full set of configurations. In Tab.~\ref{tempsetupclean} the information of Tab.~\ref{tempsetup} is repeated, but including the number of configurations that survive the selection procedure.

\begin{figure*}[t]
   \begin{center}
      \begin{subfigure}{\columnwidth}
       \includegraphics[scale=0.25,angle=-90]{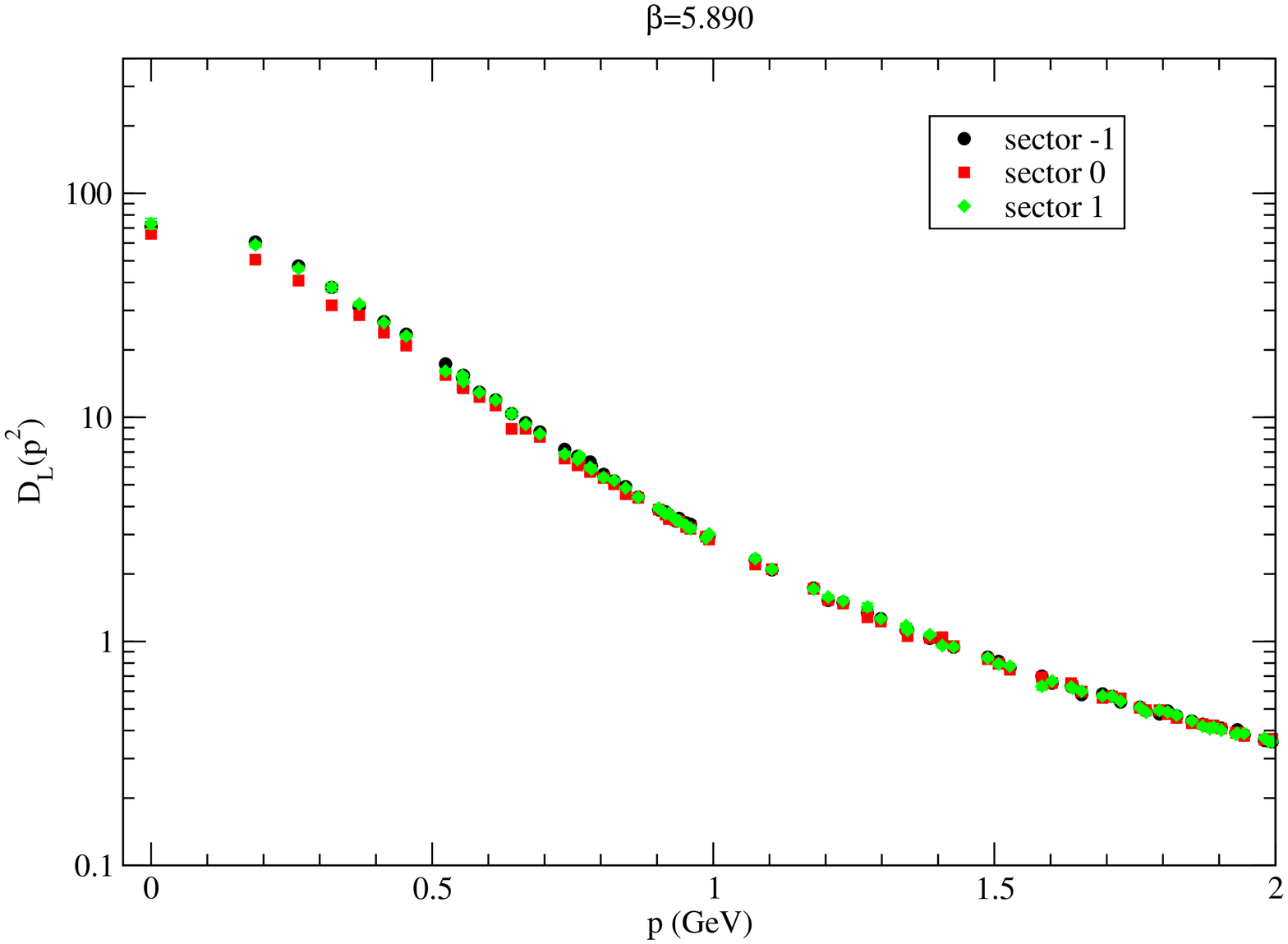} 
                        \caption{$T = 265.9$ MeV}
      \end{subfigure} \hfill
     \begin{subfigure}{\columnwidth}
      \includegraphics[scale=0.25,angle=-90]{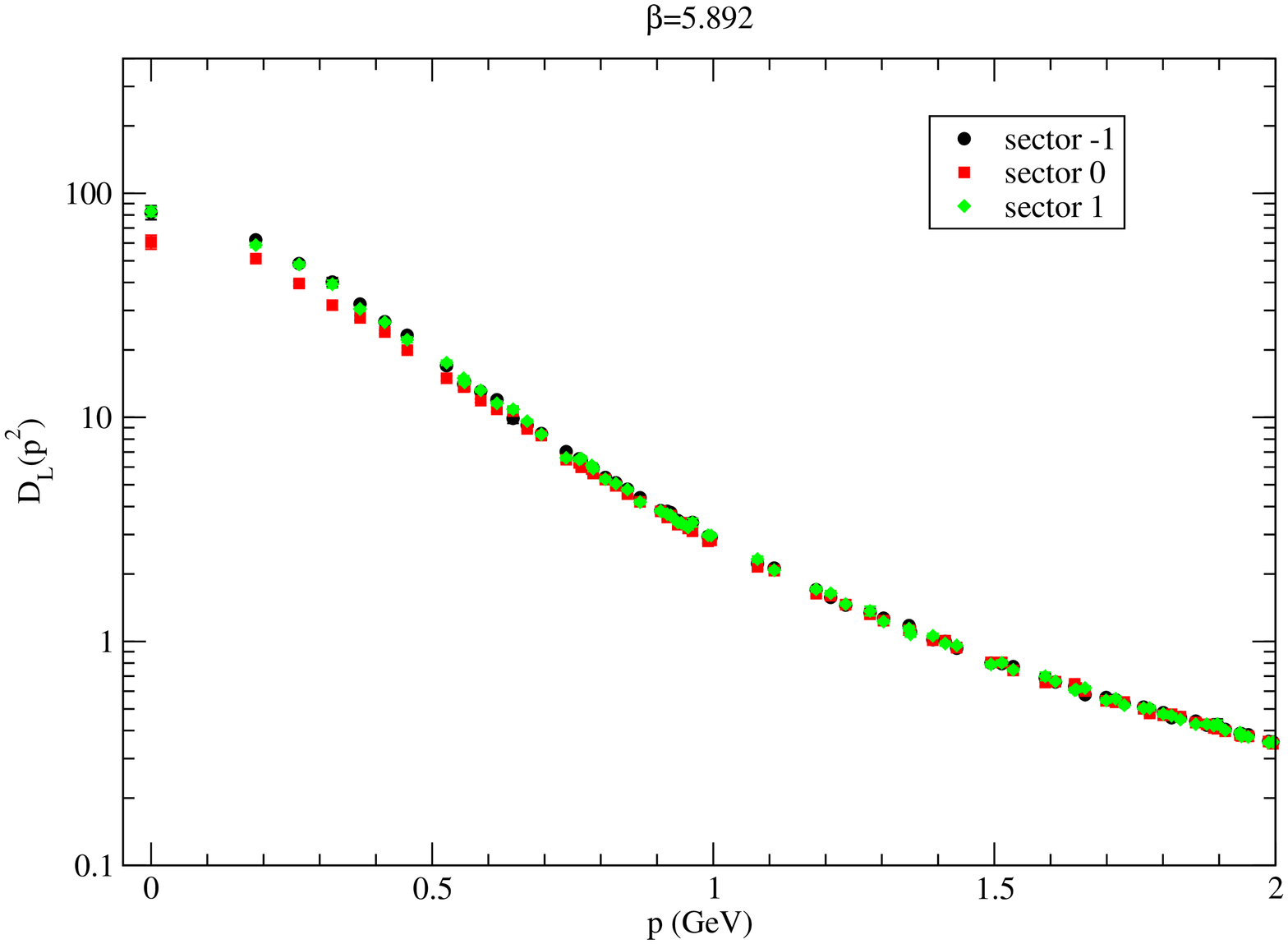} 
                       \caption{$T = 266.9$ MeV}
                       \end{subfigure} \\
     \begin{subfigure}{\columnwidth}
      \includegraphics[scale=0.25,angle=-90]{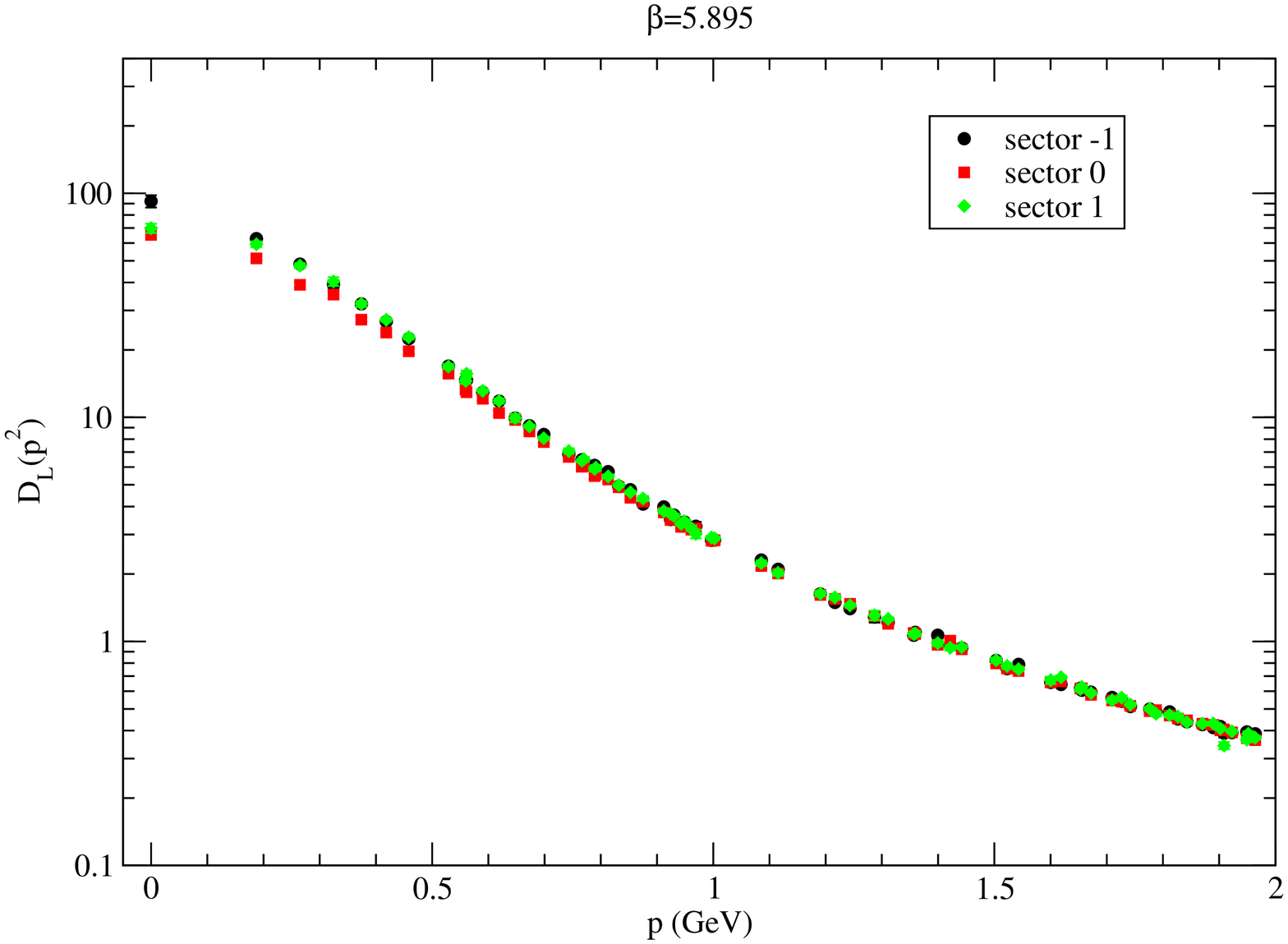} 
                       \caption{$T = 268.5$ MeV} 
                       \end{subfigure}  \hfill
      \begin{subfigure}{\columnwidth}
       \includegraphics[scale=0.25,angle=-90]{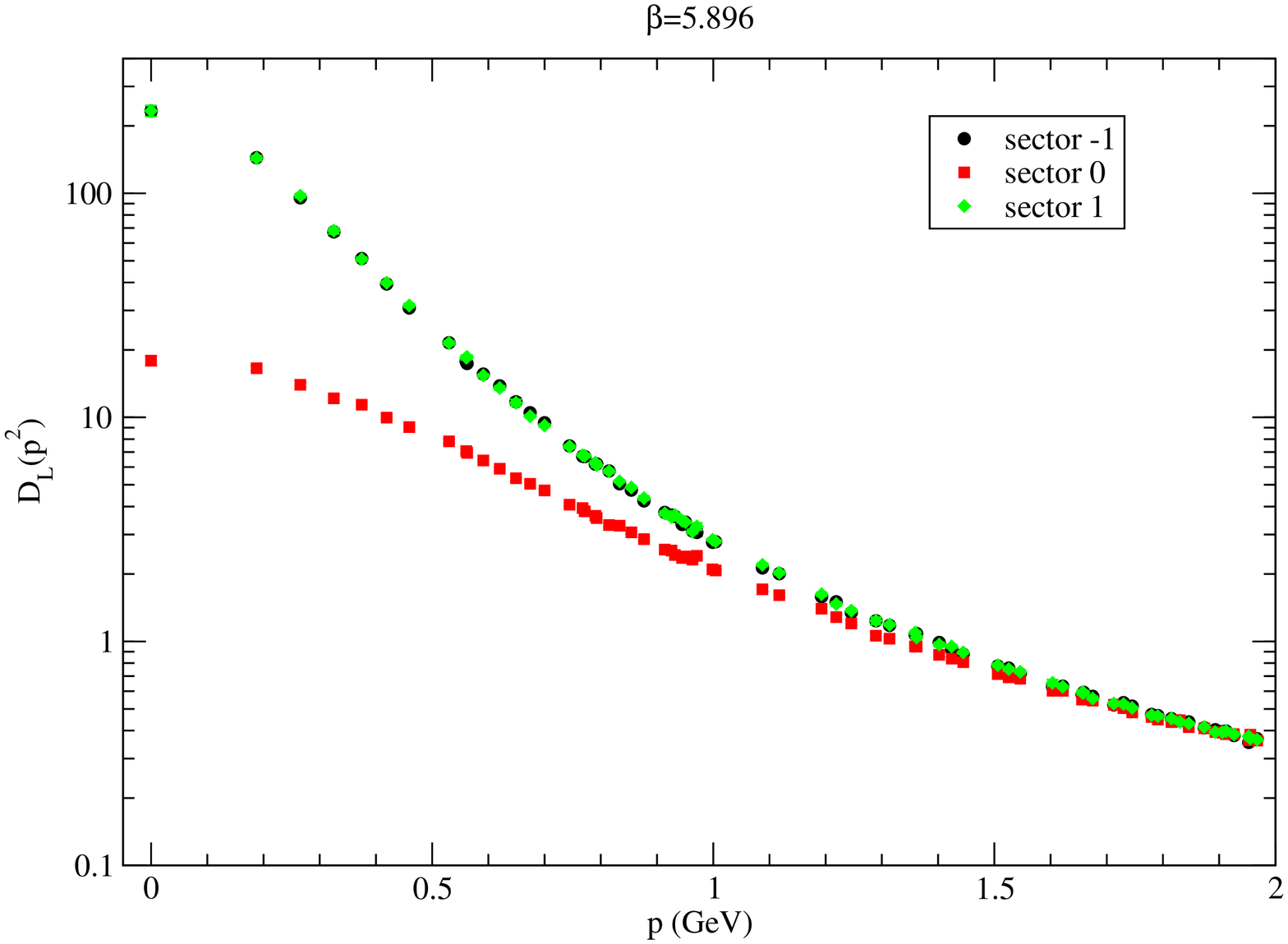} 
                        \caption{$T = 269.0$ MeV}
                        \end{subfigure} \\
      \begin{subfigure}{\columnwidth}
       \includegraphics[scale=0.25,angle=-90]{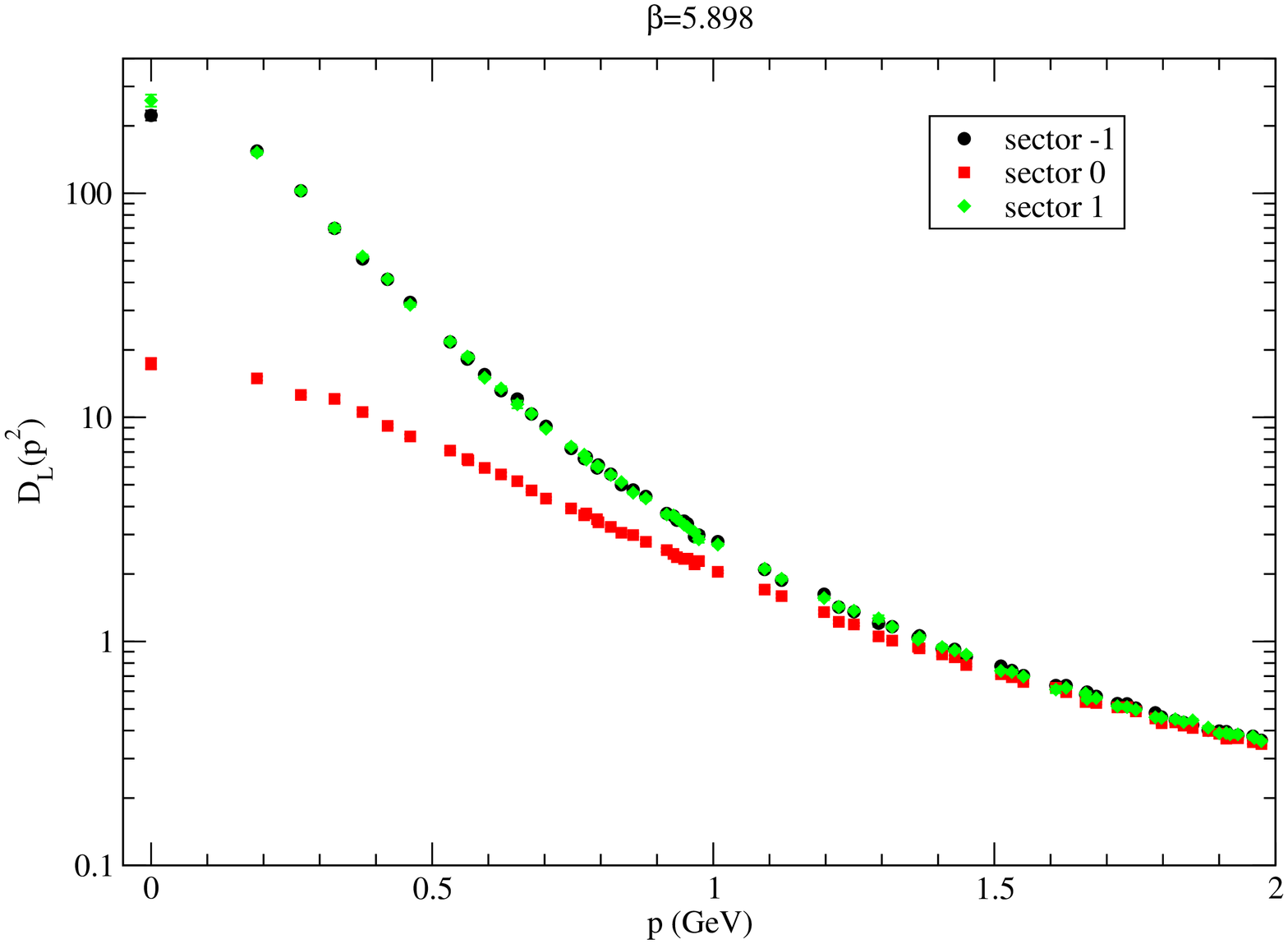} 
                        \caption{$T = 270.0$ MeV}
                        \end{subfigure} \hfill
      \begin{subfigure}{\columnwidth}
       \includegraphics[scale=0.25,angle=-90]{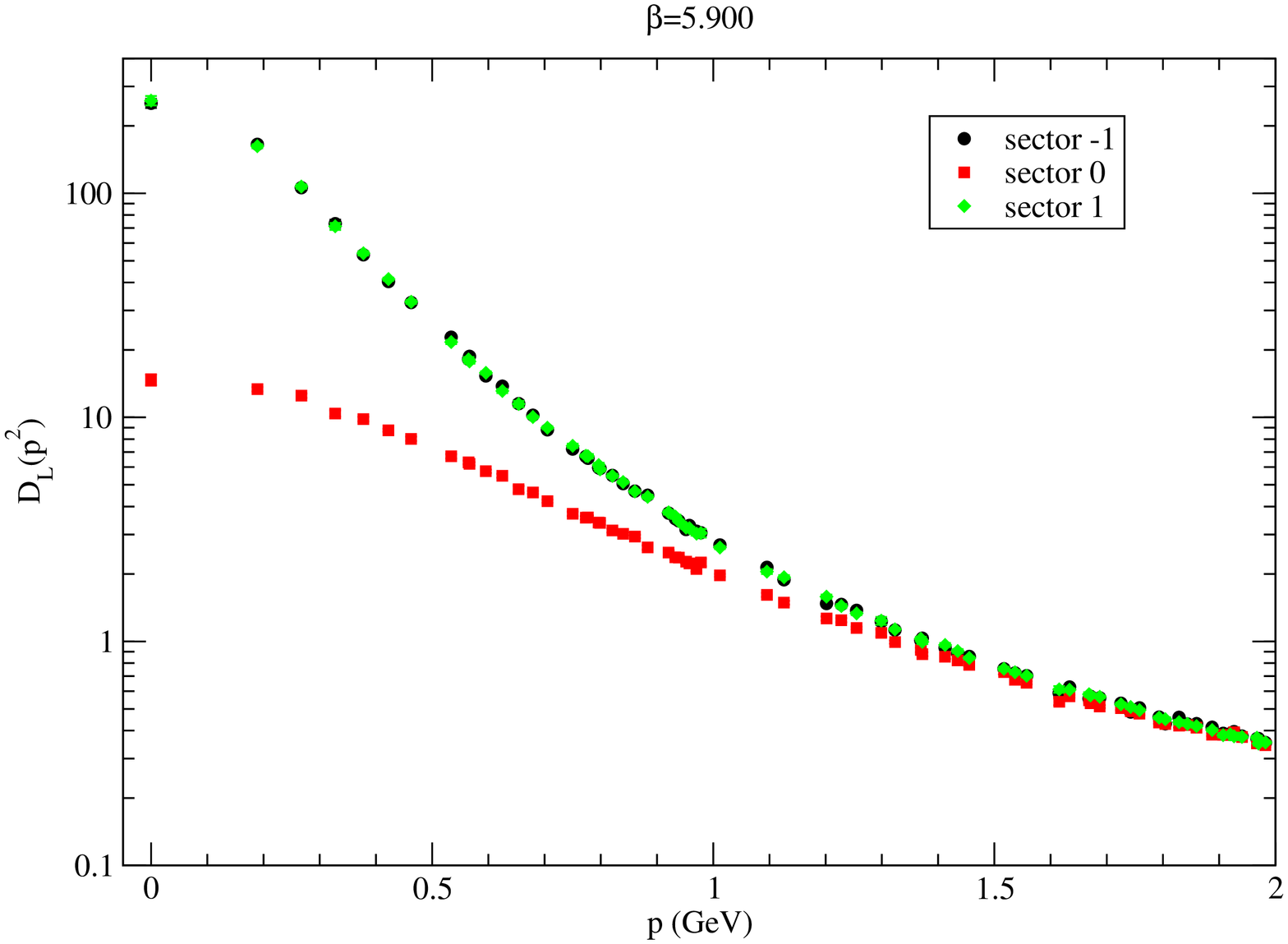} 
                        \caption{$T = 271.0$ MeV} 
                        \end{subfigure}
   \end{center}
    \caption{Electric gluon form factor $D_L(p^2,T)$ for simulations using the coarser $54^3 \times 6$ lattices.}
    \label{fig:DL2_T_54}
\end{figure*}

\begin{figure*}[t]
   \begin{center}
      \begin{subfigure}{\columnwidth}
       \includegraphics[scale=0.25,angle=-90]{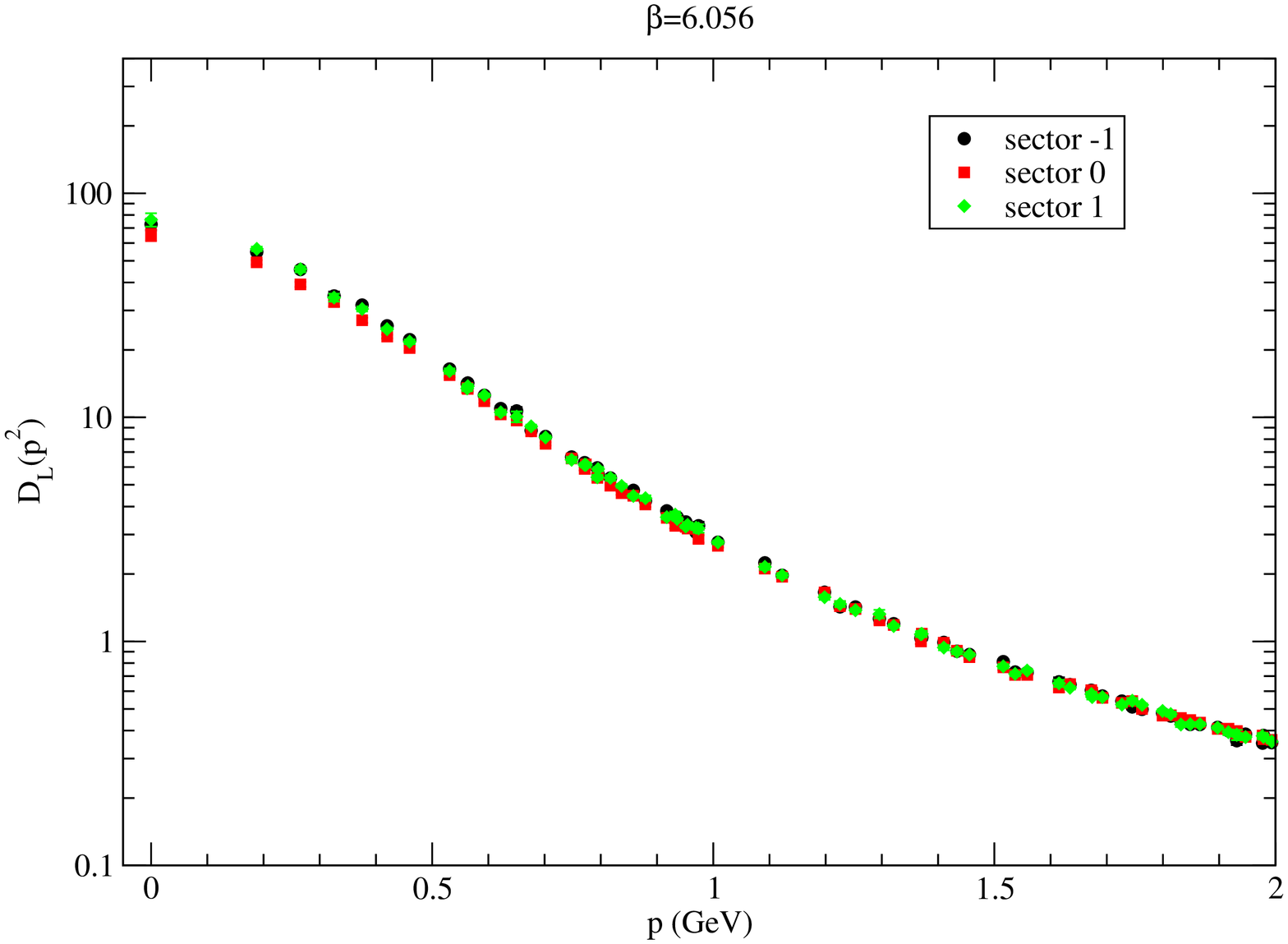} 
                        \caption{$T = 269.2$ MeV}
      \end{subfigure} \hfill
      \begin{subfigure}{\columnwidth}
      \includegraphics[scale=0.25,angle=-90]{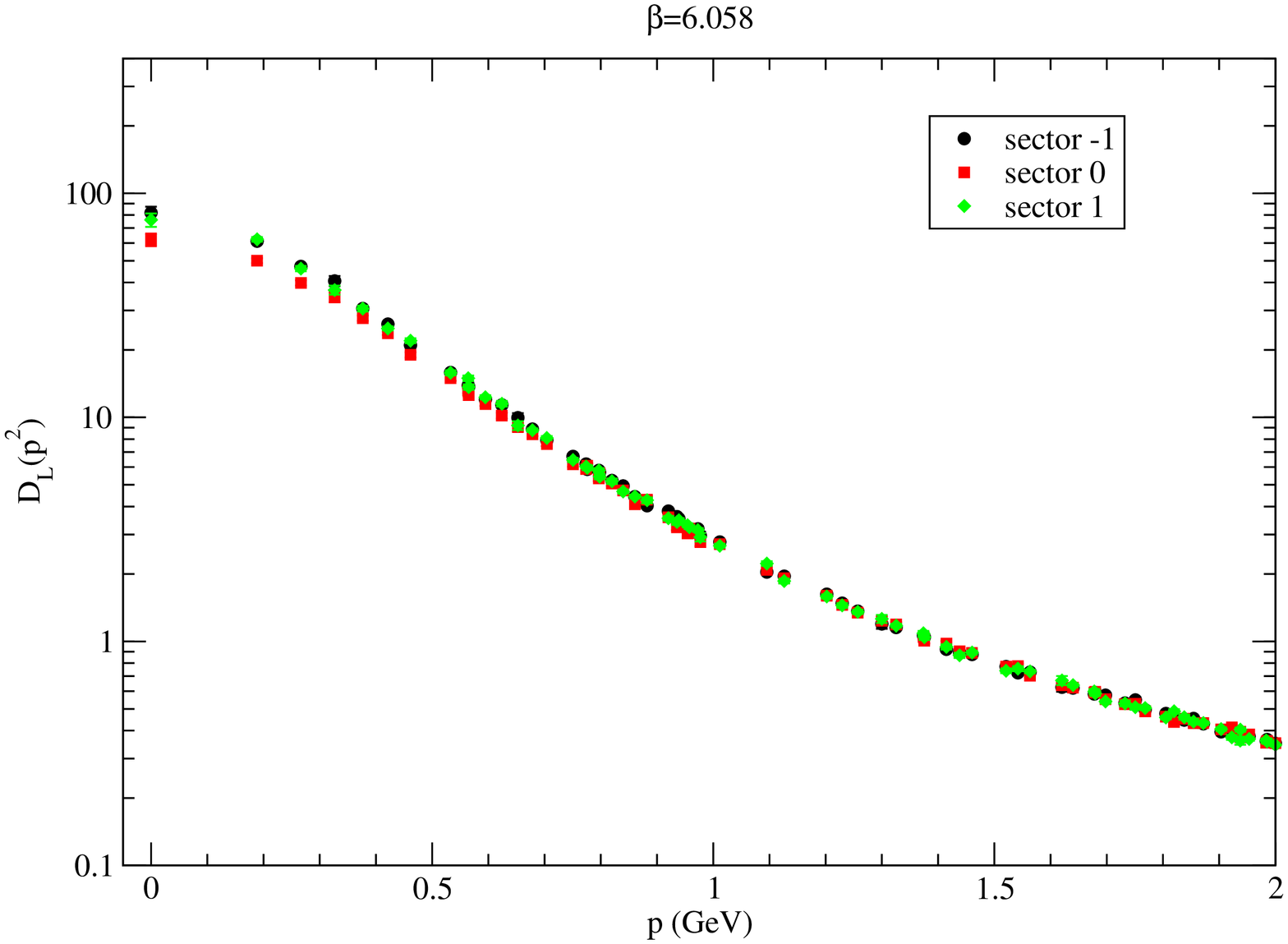} 
                       \caption{$T = 270.1$ MeV} 
                       \end{subfigure} \\
     \begin{subfigure}{\columnwidth}
      \includegraphics[scale=0.25,angle=-90]{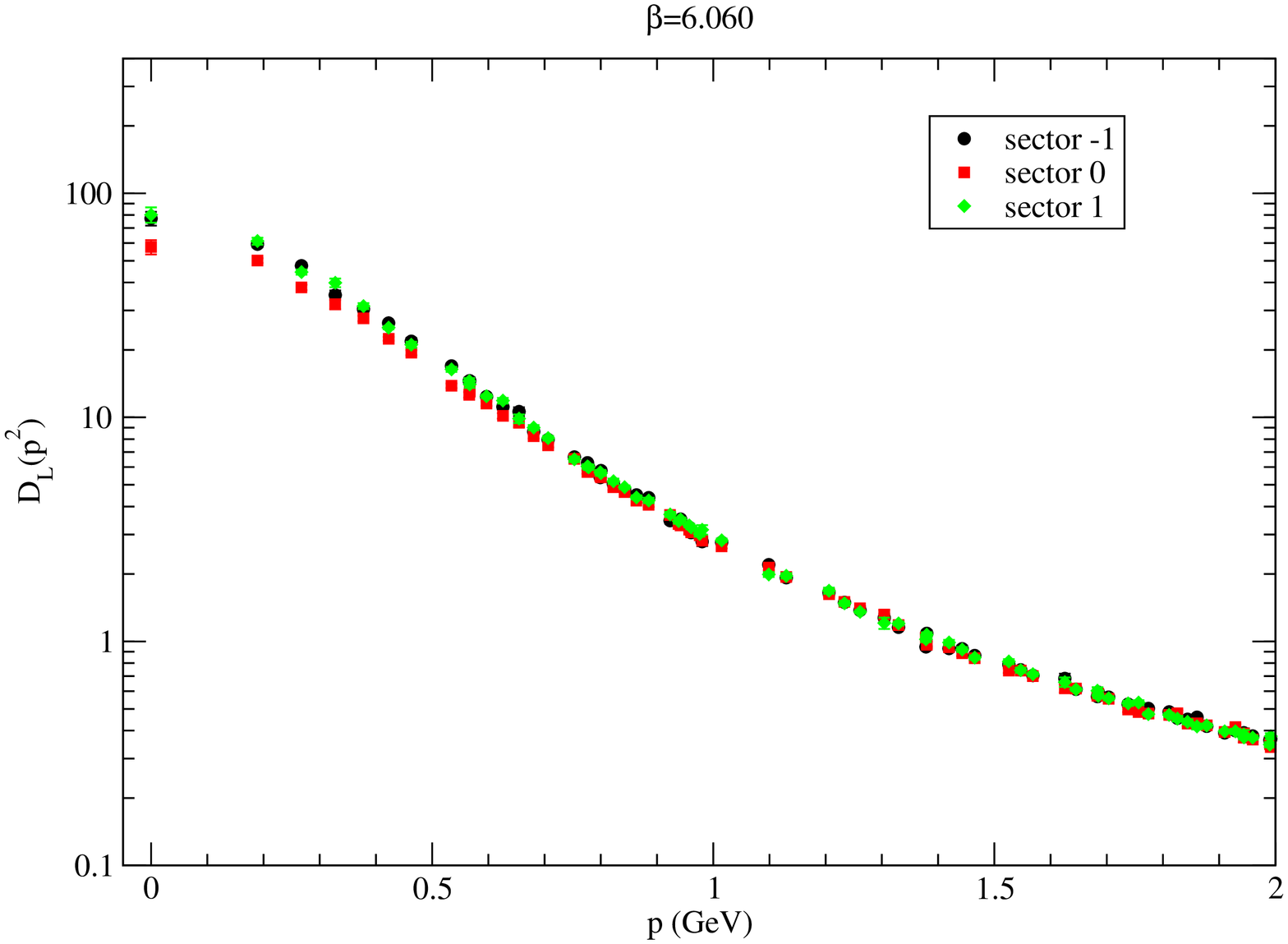} 
                       \caption{$T = 271.0$ MeV}
                       \end{subfigure} \hfill
     \begin{subfigure}{\columnwidth}
      \includegraphics[scale=0.25,angle=-90]{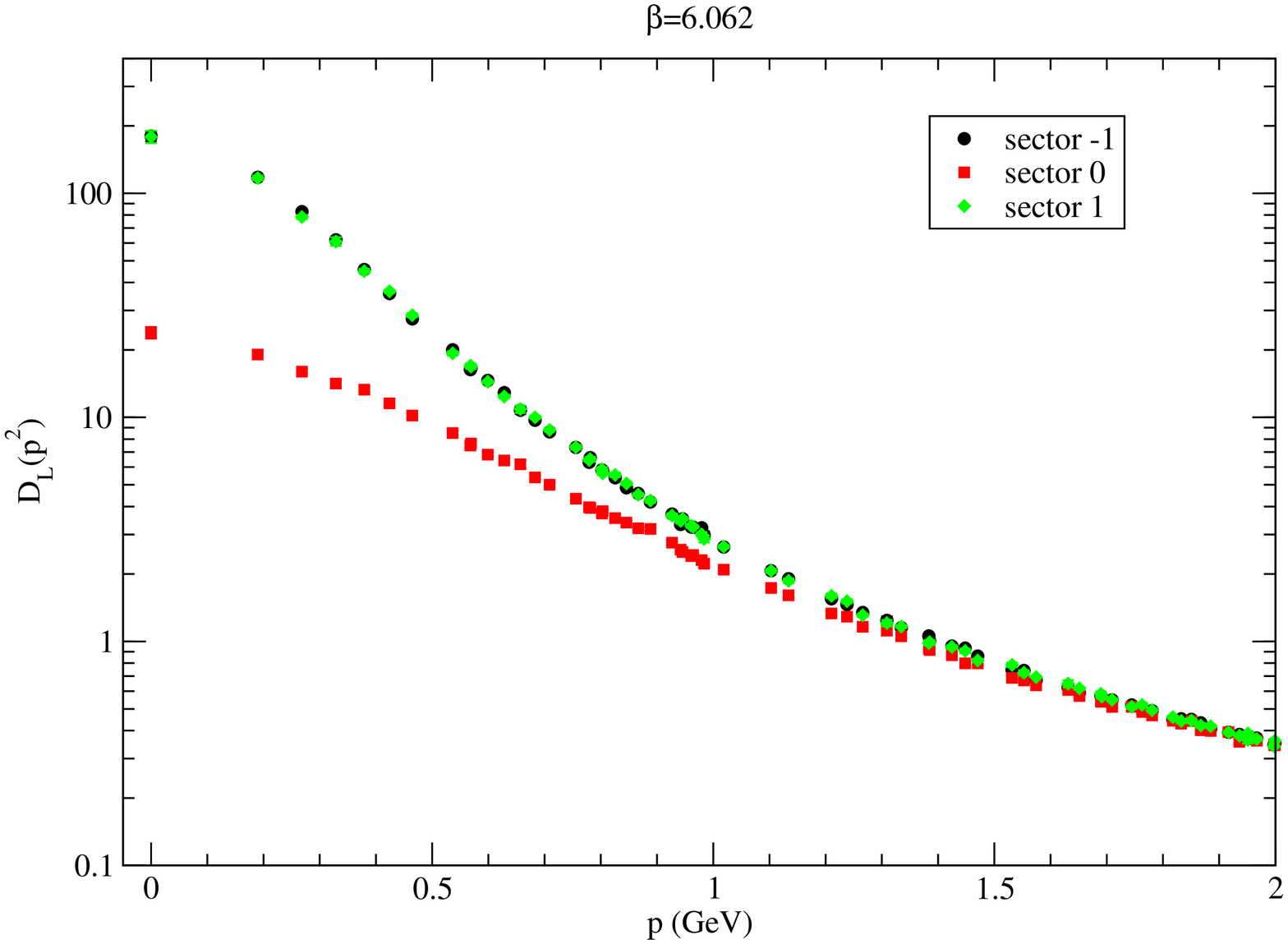} 
                       \caption{$T = 271.9$ MeV}
                       \end{subfigure} \hfill
     \begin{subfigure}{\columnwidth}
      \includegraphics[scale=0.25,angle=-90]{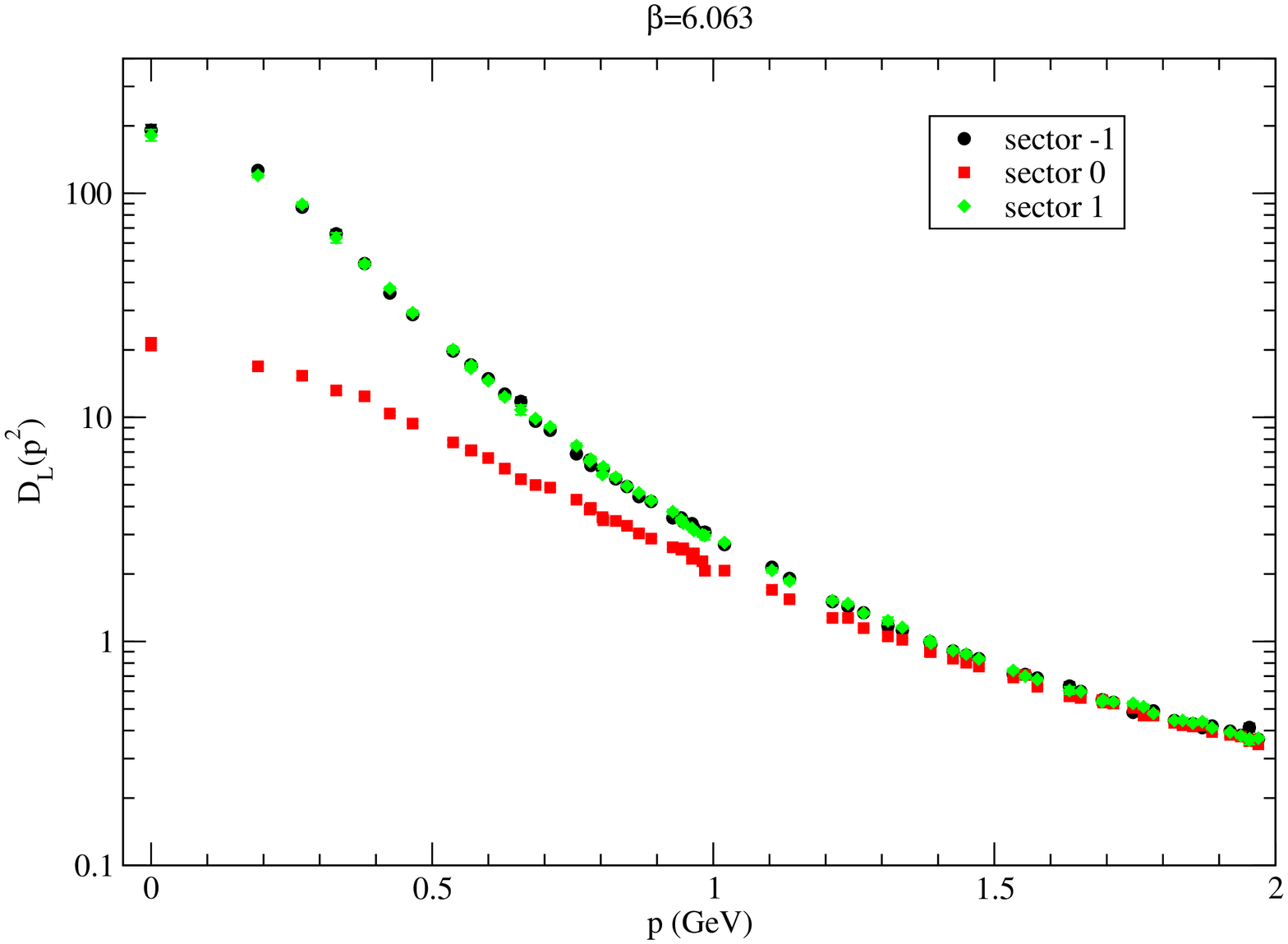} 
                       \caption{$T = 272.4$ MeV} 
                       \end{subfigure} \hfill
      \begin{subfigure}{\columnwidth}
       \includegraphics[scale=0.25,angle=-90]{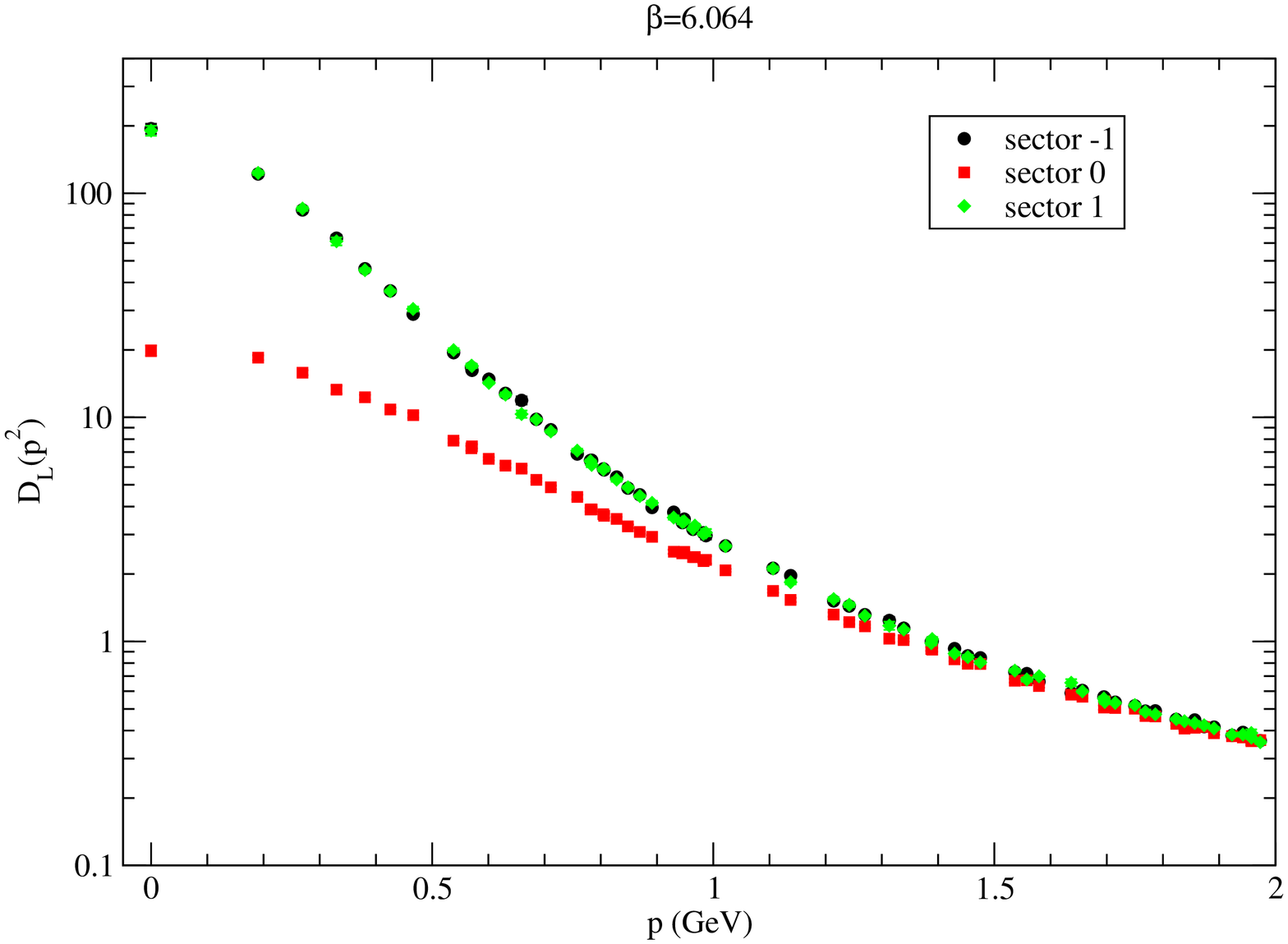} 
                        \caption{$T = 272.9$ MeV}
      \end{subfigure} 
  \end{center}
   \caption{Electric gluon form factor $D_L(p^2,T)$ for simulations using the finer $72^3 \times 8$ lattices.}
   \label{fig:DL2_T_72}
\end{figure*}

\begin{figure*}[t]
   \begin{center}
      \begin{subfigure}{\columnwidth}
       \includegraphics[scale=0.25,angle=-90]{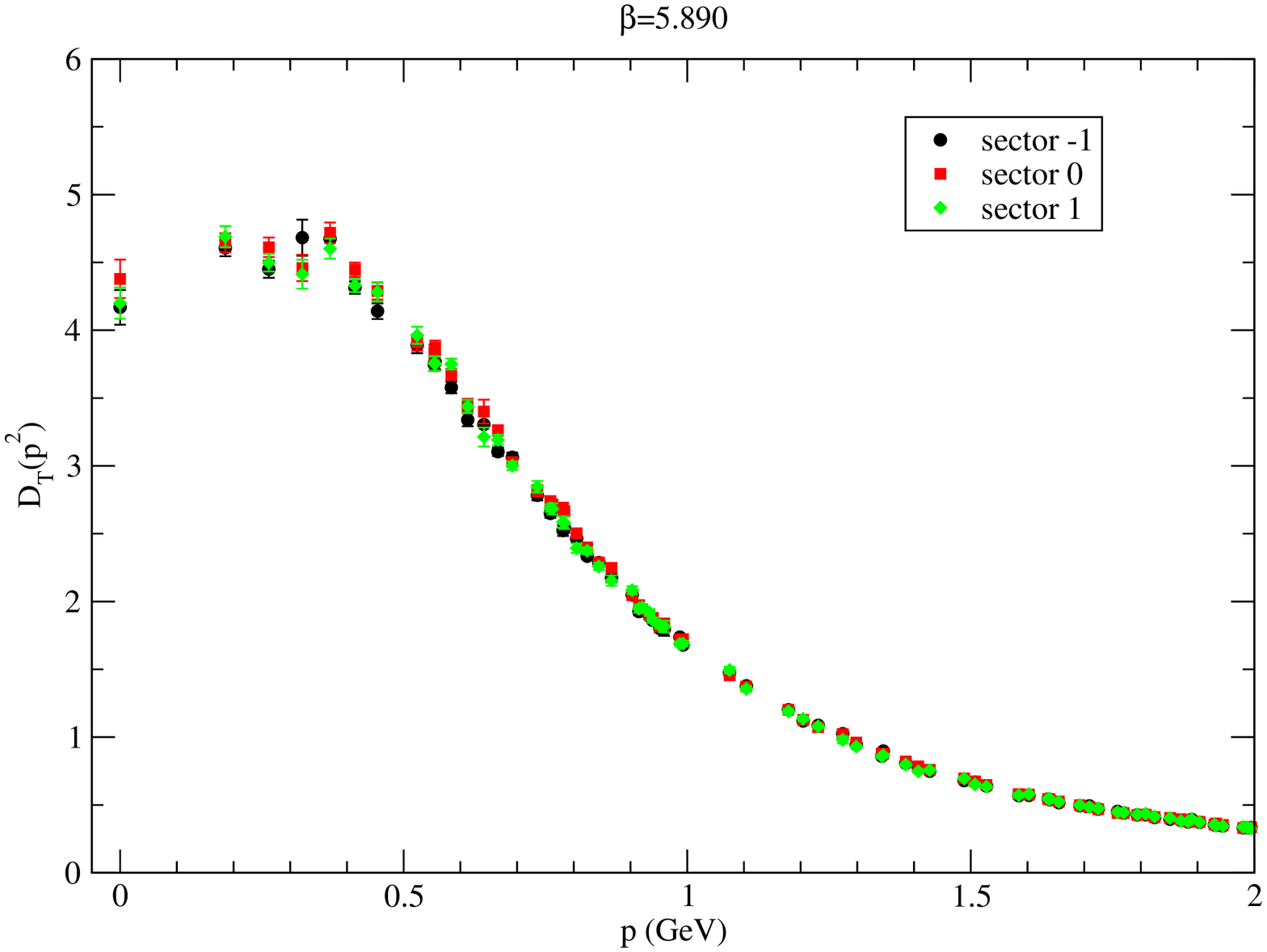} 
                        \caption{$T = 265.9$ MeV}
      \end{subfigure} \hfill
     \begin{subfigure}{\columnwidth}
      \includegraphics[scale=0.25,angle=-90]{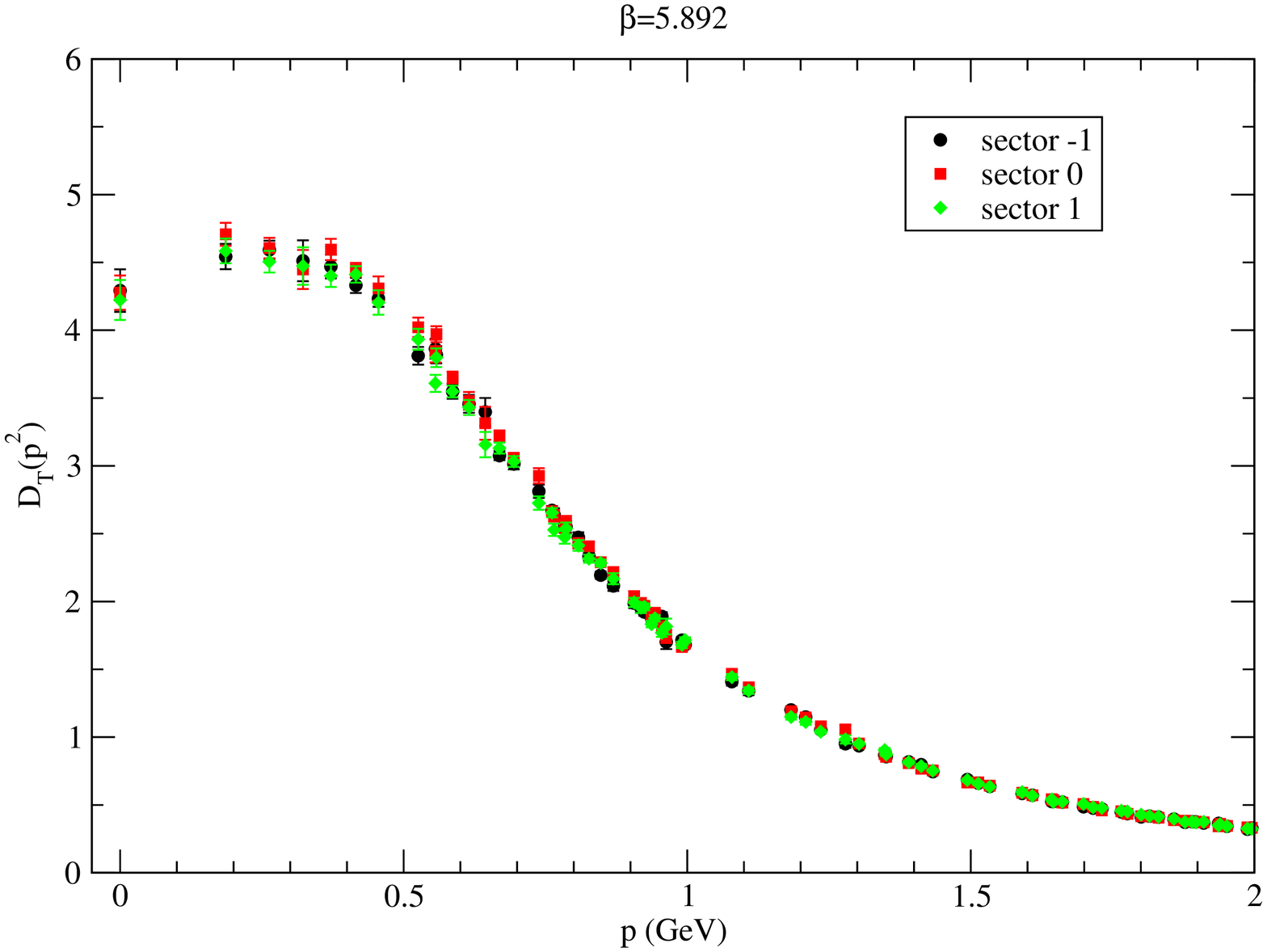} 
                       \caption{$T = 266.9$ MeV}
                       \end{subfigure} \\
     \begin{subfigure}{\columnwidth}
      \includegraphics[scale=0.25,angle=-90]{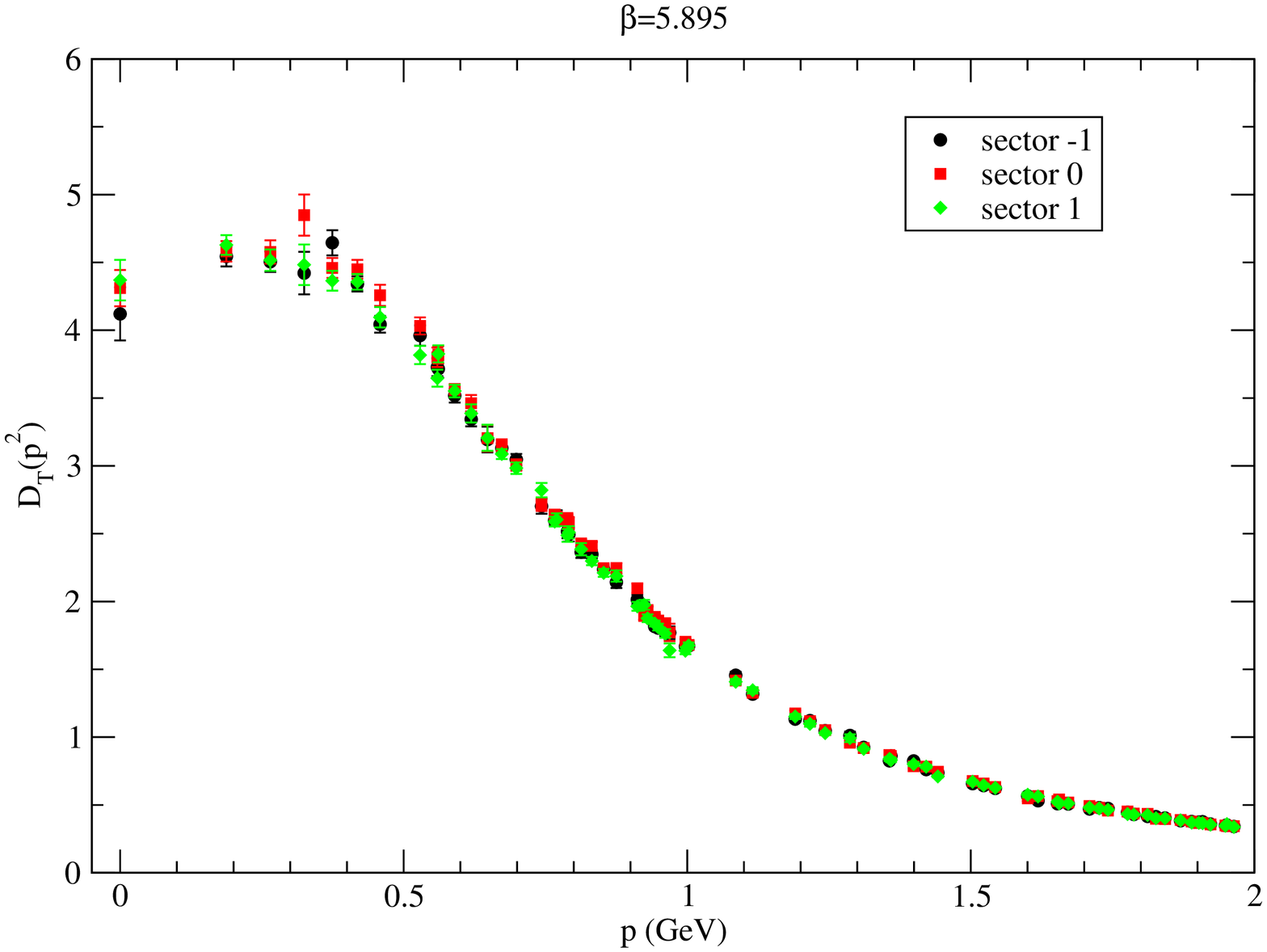} 
                       \caption{$T = 268.5$ MeV} 
                       \end{subfigure} 
      \begin{subfigure}{\columnwidth}
       \includegraphics[scale=0.25,angle=-90]{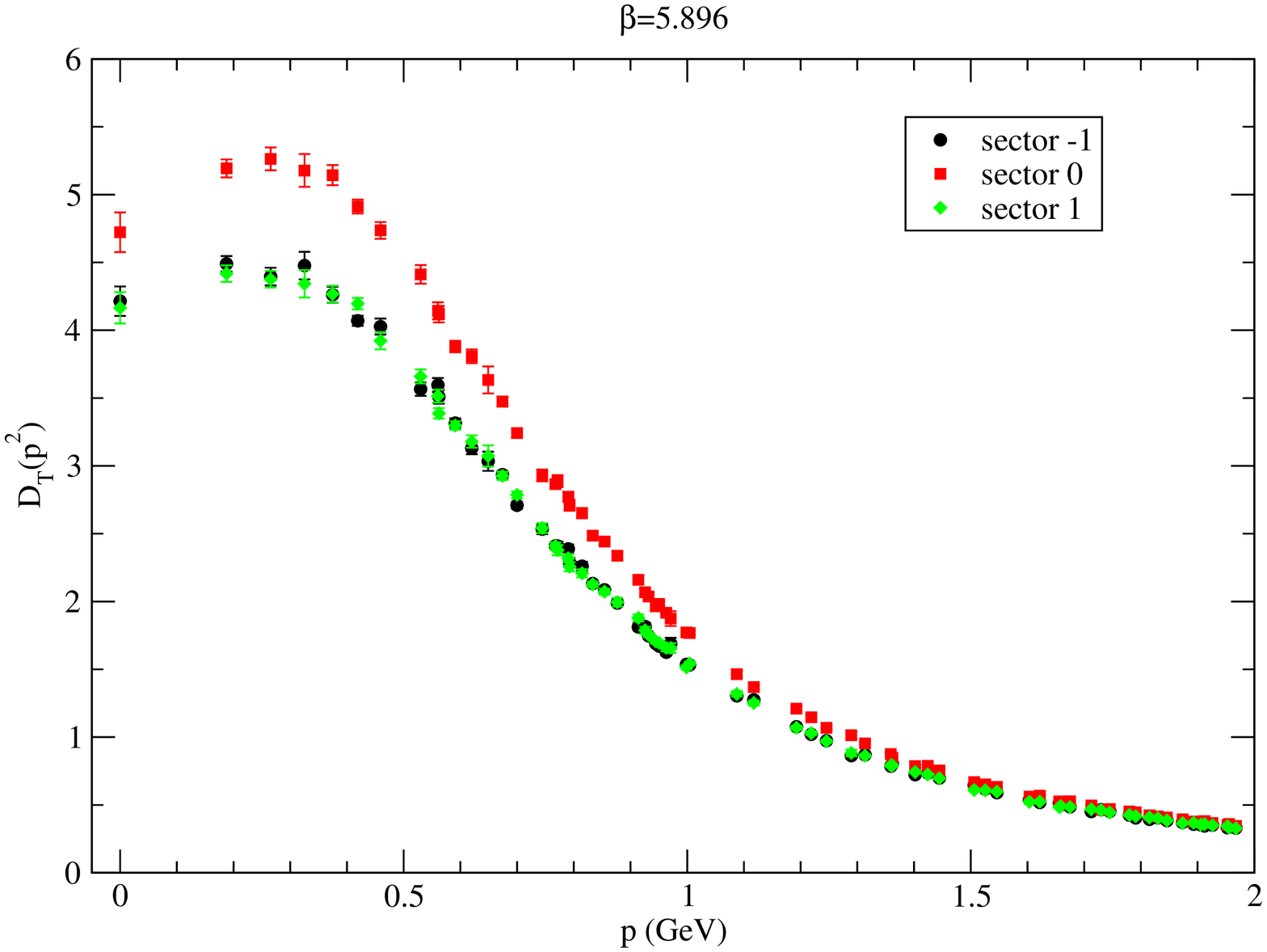} 
                        \caption{$T = 269.0$ MeV}
                        \end{subfigure} \\
      \begin{subfigure}{\columnwidth}
       \includegraphics[scale=0.25,angle=-90]{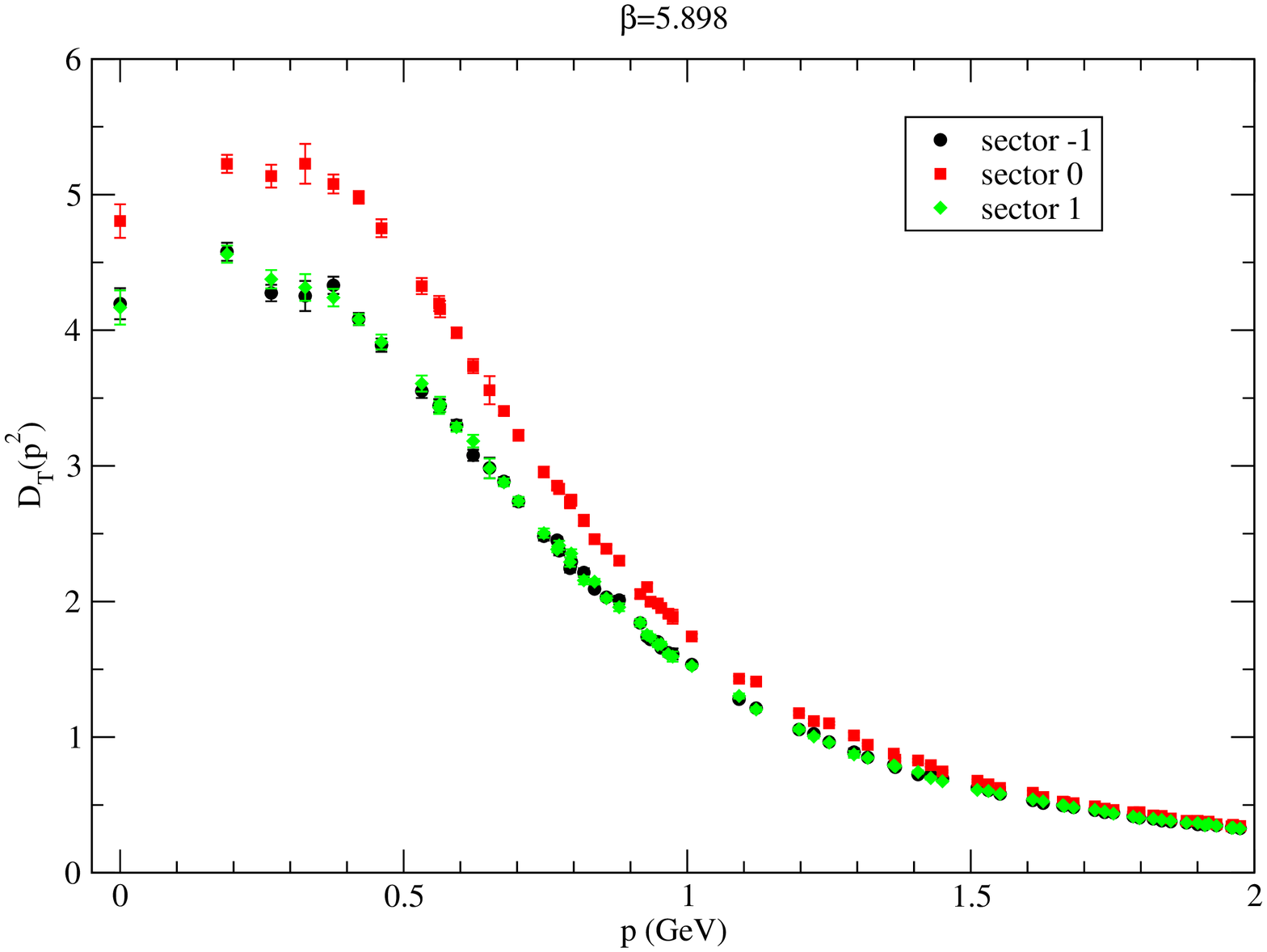} 
                        \caption{$T = 270.0$ MeV}
                        \end{subfigure} \hfill
      \begin{subfigure}{\columnwidth}
       \includegraphics[scale=0.25,angle=-90]{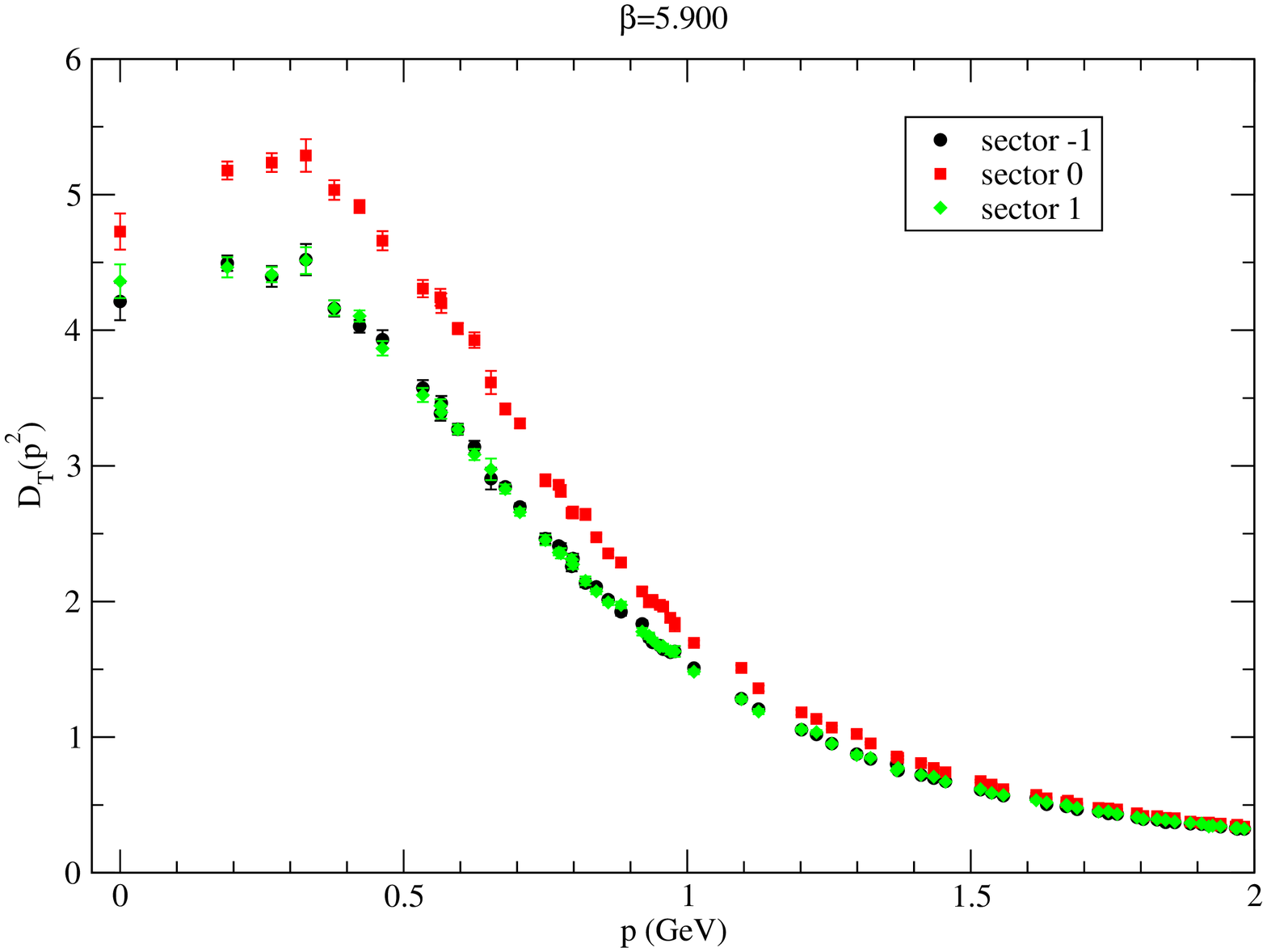} 
                        \caption{$T = 271.0$ MeV} 
                        \end{subfigure}
   \end{center}
    \caption{Magnetic gluon form factor $D_T(p^2,T)$ for simulations using the coarser $54^3 \times 6$ lattices.}
    \label{fig:DT2_T_54}
\end{figure*}

\begin{figure*}[t]
   \begin{center}
      \begin{subfigure}{\columnwidth}
       \includegraphics[scale=0.25,angle=-90]{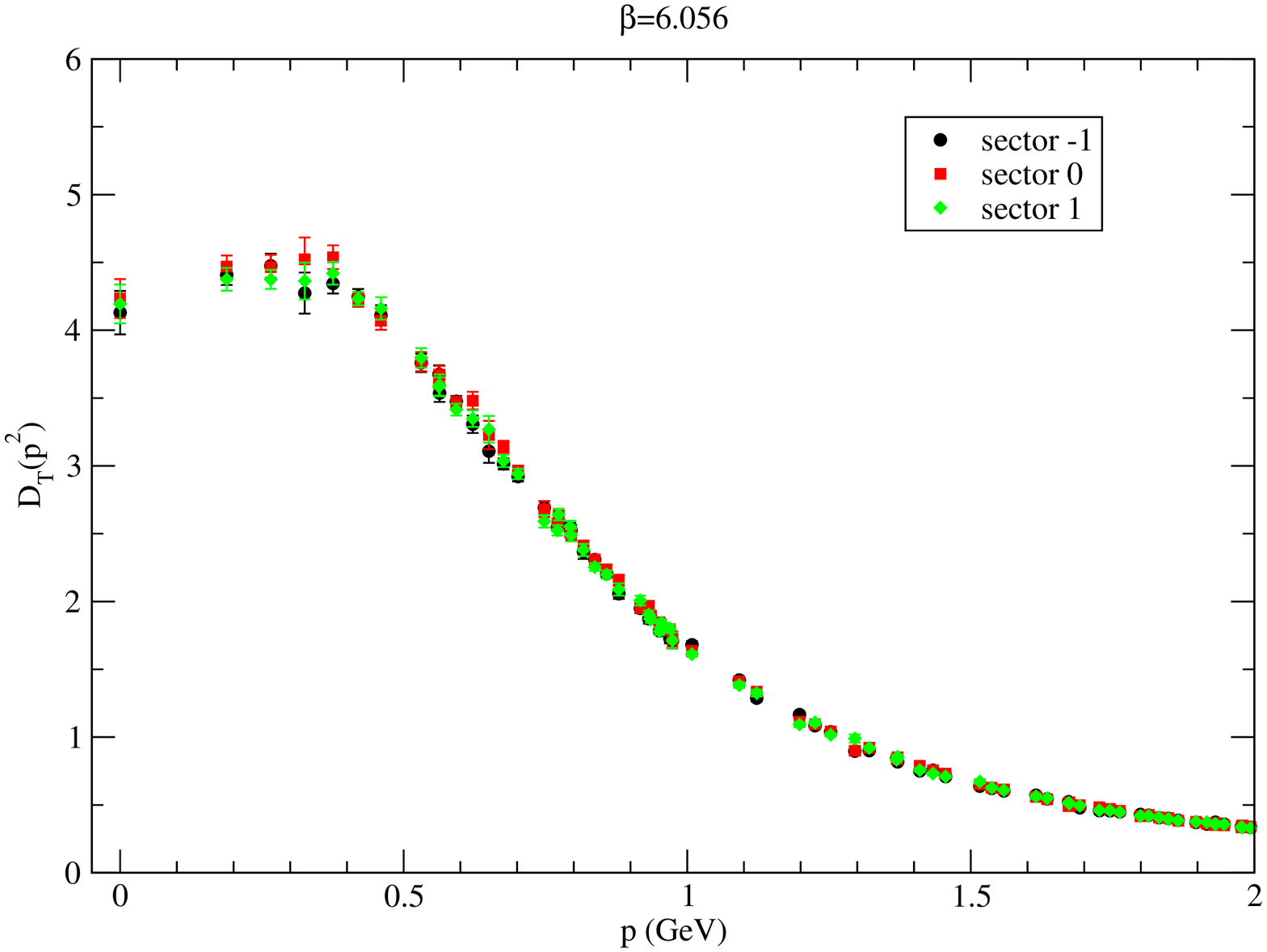} 
                        \caption{$T = 269.2$ MeV}
      \end{subfigure} \hfill
      \begin{subfigure}{\columnwidth}
      \includegraphics[scale=0.25,angle=-90]{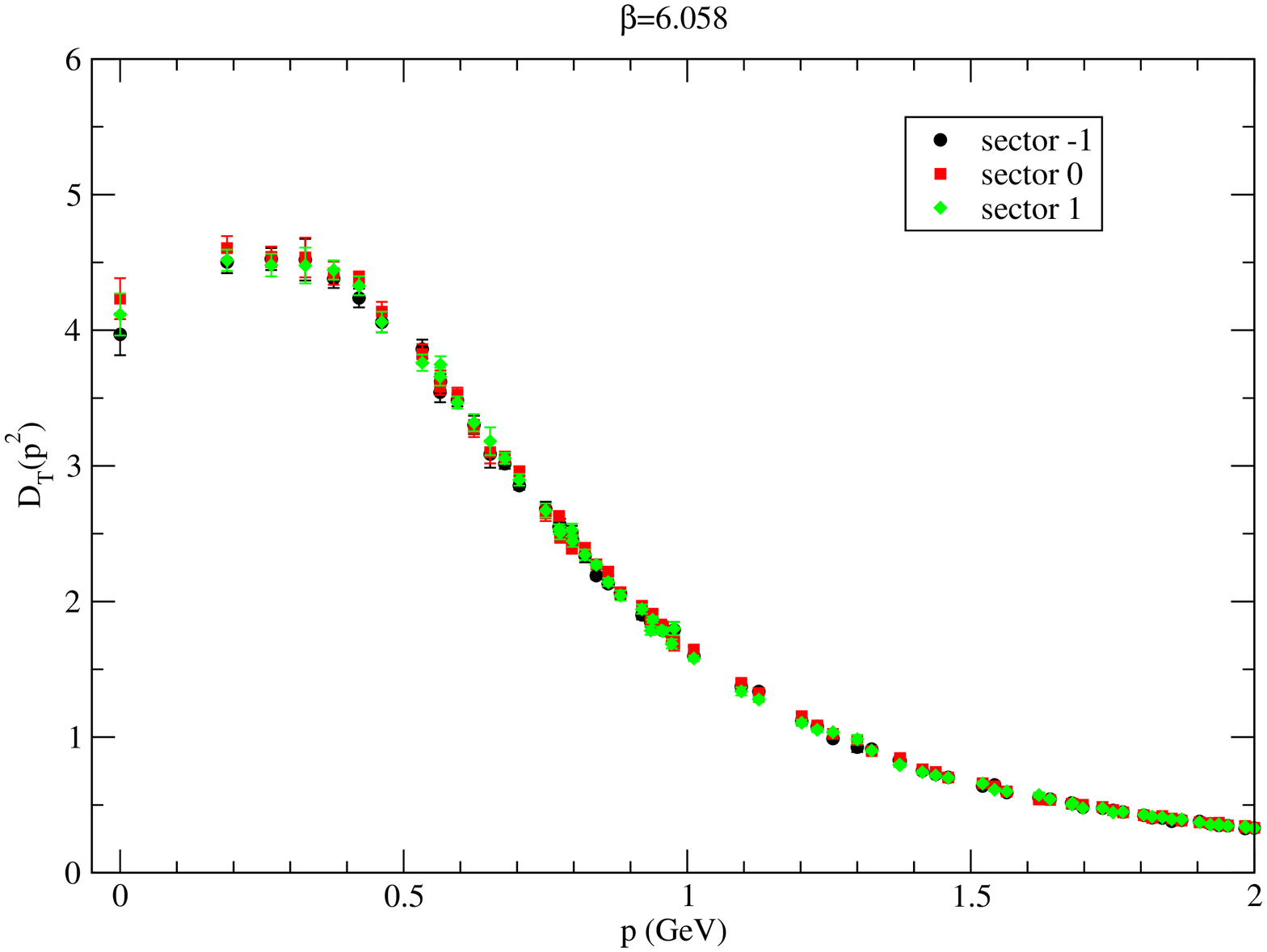} 
                       \caption{$T = 270.1$ MeV} 
                       \end{subfigure} \\
     \begin{subfigure}{\columnwidth}
      \includegraphics[scale=0.25,angle=-90]{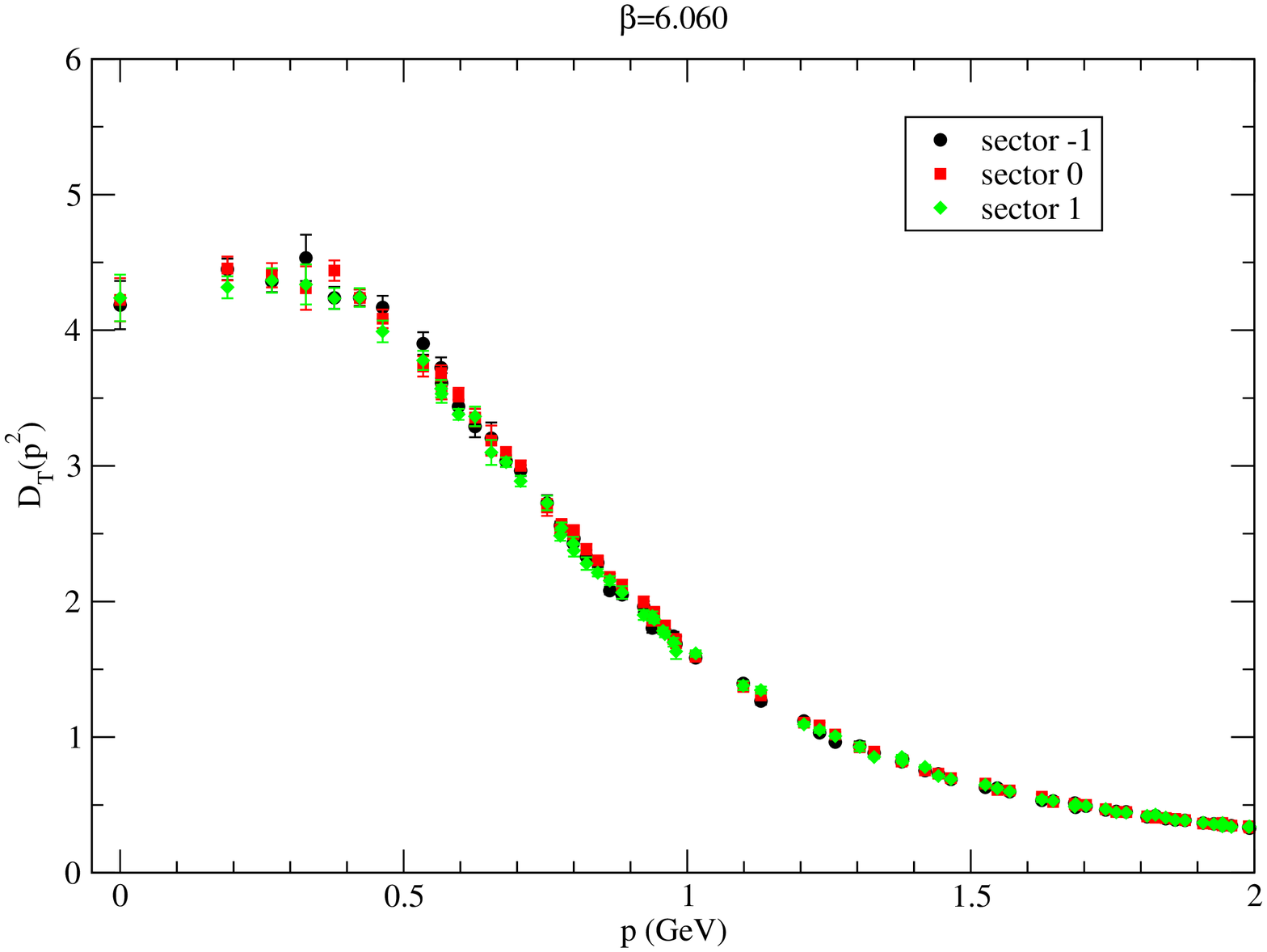} 
                       \caption{$T = 271.0$ MeV}
                       \end{subfigure} \hfill
     \begin{subfigure}{\columnwidth}
      \includegraphics[scale=0.25,angle=-90]{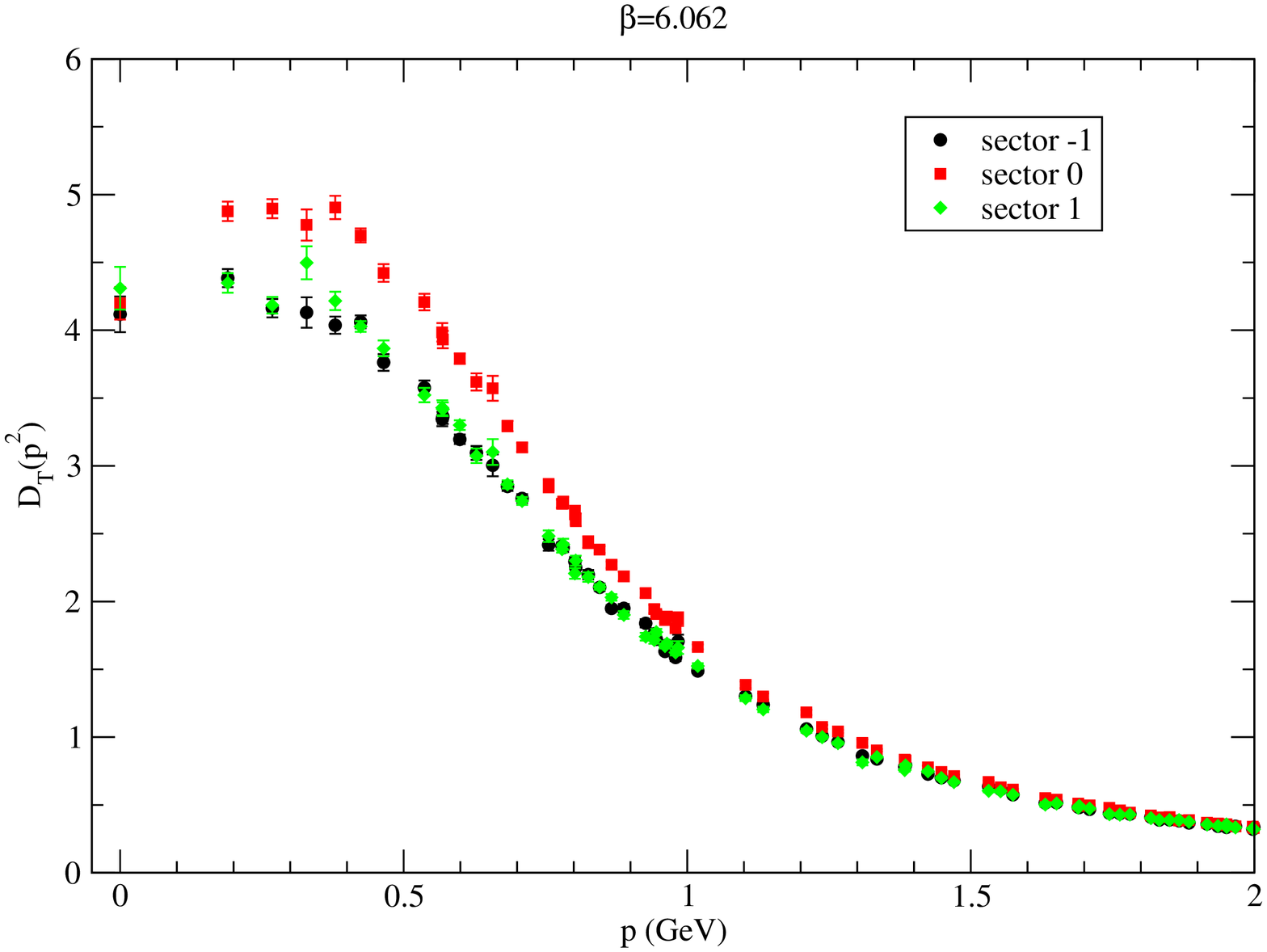} 
                       \caption{$T = 271.9$ MeV}
                       \end{subfigure} \\
     \begin{subfigure}{\columnwidth}
      \includegraphics[scale=0.25,angle=-90]{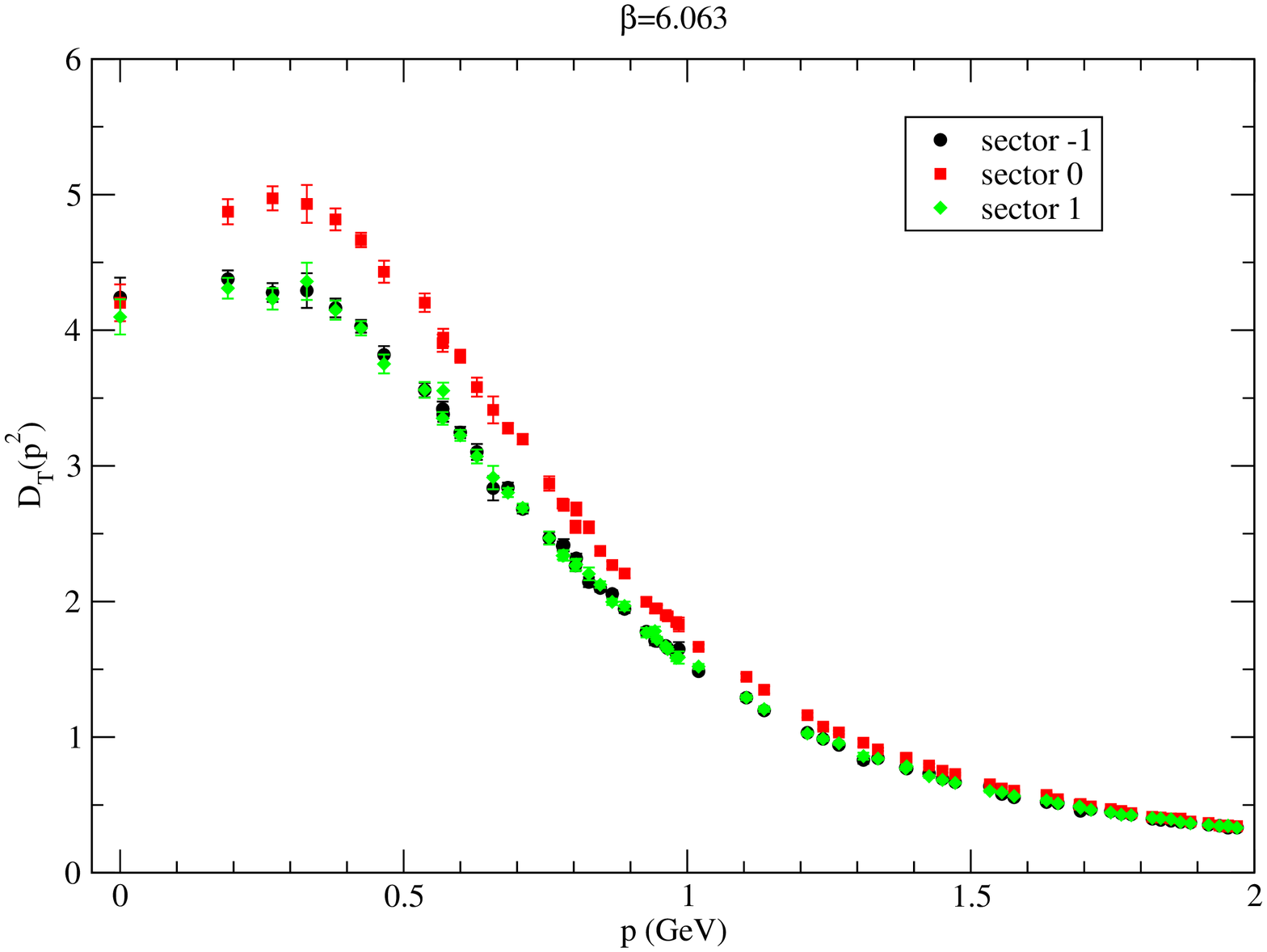} 
                       \caption{$T = 272.4$ MeV} 
                       \end{subfigure}  \hfill
      \begin{subfigure}{\columnwidth}
       \includegraphics[scale=0.25,angle=-90]{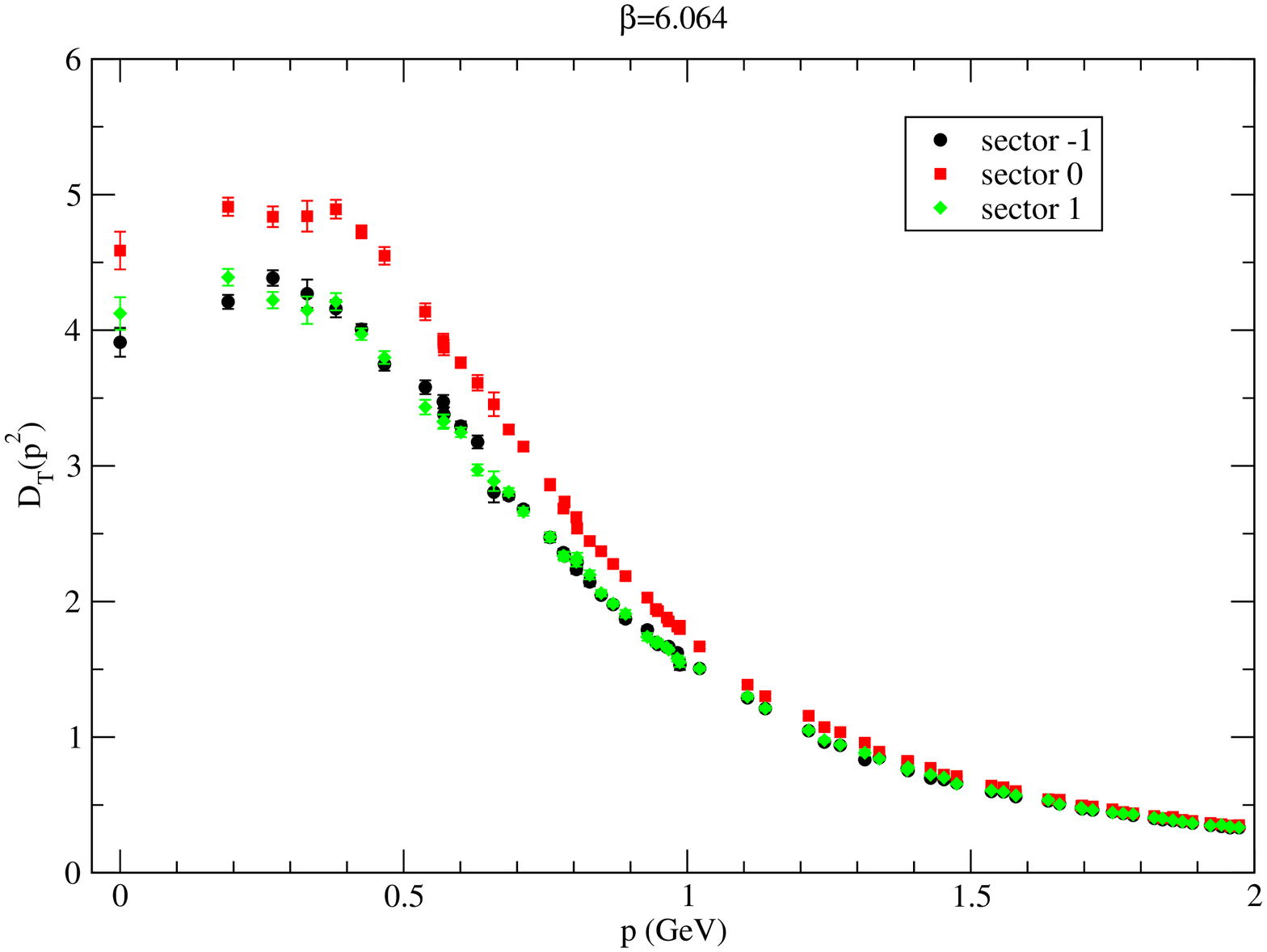} 
                        \caption{$T = 272.9$ MeV}
      \end{subfigure}
  \end{center}
   \caption{Magnetic gluon form factor $D_T(p^2,T)$ for simulations using the finer $72^3 \times 8$ lattices.}
   \label{fig:DT2_T_72}
\end{figure*}

In Figs. \ref{fig:DL2_T_54}, \ref{fig:DL2_T_72}, \ref{fig:DT2_T_54} and \ref{fig:DT2_T_72} we report on  
the two gluon form factors
 for the temperatures reported in Tab.~\ref{tempsetup} and after applying the selection procedure to the gauge configurations. 
As the figures show, it is in the electric sector where the deconfinement transition has a larger impact. Indeed, for the zero sector $D_L(p^2;T)$
is strongly suppressed in the deconfined phase, while the magnetic sector has a much more modest increase as T crosses $T_c$.
This reproduces the behaviour already observed in various simulations; see, e.g., \cite{SilvaOliveiraetal2014,Aouane12} and references there in.

Moreover, for $T > T_c$ the gluon form factors associated to the various $Z_3$ sectors are not only different quantitatively but also qualitatively. 
This difference is observed only between the 0 sector and the $\pm 1$ sectors. 
The longitudinal propagator $D_L$ associated with the  $\pm 1$ sectors is strongly enhanced if compared with the zero sector result. In fact, for the 
 zero sector, $D_L$ defines a larger mass scale,
compared to the corresponding longitudinal propagators associated with the others sectors. 
On the other hand, the transverse propagator $D_T$ for the 0 sector seems to take higher values, and therefore one can associate a smaller mass scale,
in comparison with transverse propagators defined in the $\pm 1$ sectors. 

The separation of $D_L$ and $D_T$ associated with the various $Z_3$ sectors starts around the deconfinement
phase transition. For the coarser lattice, this separation starts to show up at $T = 267$ MeV for $D_L$ and it is clearly
seen for $T = 269$ MeV and above. For the magnetic form factor $D_T$, the difference between the sectors sets in at $T = 269$ MeV.
For the finer lattice, the differences in $D_L$ for the various $Z_3$ sectors start at $T=270$ MeV, while $D_T$ distinguishes
the various sectors for $T = 272$ MeV and above. 
From the separation of the gluon form factors one can identify a deconfinement phase transition at $T_c = 267 - 272$ MeV, in agreement with 
the  values quoted in the literature for the pure SU(3) gauge theory.

%+++++++++++++++++++++++++++++++++++++++++++++++++++++++++++++++++++++++++++
%+++++++++++++++++++++++++++++++++++++++++++++++++++++++++++++++++++++++++++
%+++++++++++++++++++++++++++++++++++++++++++++++++++++++++++++++++++++++++++
\begin{table}[t!]
\begin{center}
\begin{tabular}{lc@{\hspace{0.5cm}}l@{\hspace{0.5cm}}l@{\hspace{0.5cm}}l}
\hline
         &            & $54^3  \times 6$ & $72^3  \times 8$     & $90^3  \times 10$  \\
         & Sector &   $\beta=5.896$  &  $\beta=6.062$       &   $\beta=6.212$    \\
 T =   &            &  $269$ MeV        &   $271.9$ MeV        &  $276.6$ MeV \\
\hline
         & $-1$          &  4.21 $\pm$ 0.11        &  4.12 $\pm$ 0.13   &    3.81 $\pm$ 0.13  \\
$D_T(0)$ & $0$    &  4.72 $\pm$ 0.15        &  4.20 $\pm$ 0.12   &    4.51 $\pm$ 0.13  \\
         & $1$            &  4.17 $\pm$ 0.13       &  4.31 $\pm$ 0.16    &   4.04 $\pm$ 0.10  \\
\hline
         & $-1$          &  233 $\pm$ 11            &  180.1 $\pm$ 9.0   &   203 $\pm$ 12     \\
$D_L(0)$ & $0$    &  17.92 $\pm$ 0.84      &   23.9 $\pm$ 1.4    &  19.3 $\pm$ 1.0    \\
         & $1$           &  233 $\pm$ 13            &   179 $\pm$ 11      &   172.9 $\pm$ 9.3   \\
\hline
$\Delta D_L(0)$ & &  215 $\pm$ 11           &   156.2 $\pm$ 9.1    &  184 $\pm$ 12    \\
\hline
\end{tabular}
\end{center}
\caption{Comparison of $D_L(0)$ and $D_T(0)$ (in MeV$^{-2}$) for various ensembles just above $T_c$.
$\Delta D_L(0)$ (errors added in quadrature) refers to the modulus of the difference of $D_L (0)$ between the sectors 0 and -1.}
\label{Dzero_Tc}
\end{table}
%+++++++++++++++++++++++++++++++++++++++++++++++++++++++++++++++++++++++++++
%+++++++++++++++++++++++++++++++++++++++++++++++++++++++++++++++++++++++++++
%+++++++++++++++++++++++++++++++++++++++++++++++++++++++++++++++++++++++++++

The observed difference between the propagators for the various $Z_3$ sectors can be better illustrated looking at how $D_L(0)$ evolves with the
temperature. Furthermore, one can understand the effects of our  selection procedure to distinguish the configurations between the different phases by studying
$D_L(0)$ as a function of the temperature. 

\begin{figure*}[t]
   \begin{center}
      \begin{subfigure}{\columnwidth}
       \includegraphics[scale=0.3,angle=0]{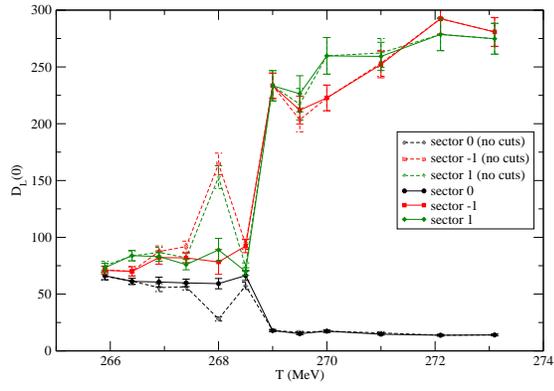} 
                        \caption{$54^3 \times 6$ lattices.}
      \end{subfigure} \hfill
      \begin{subfigure}{\columnwidth}
      \includegraphics[scale=0.3,angle=0]{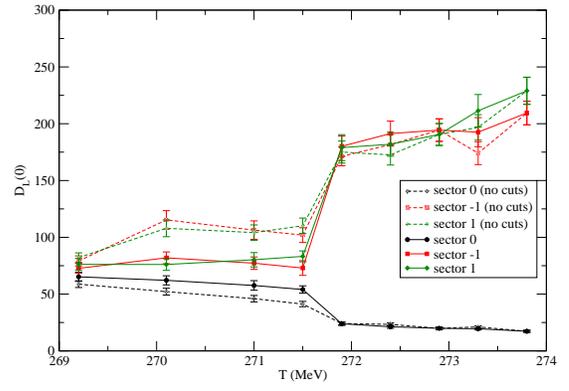} 
                       \caption{$72^3 \times 8$ lattices.} 
                       \end{subfigure} 
  \end{center}
   \caption{$D_L(0)$ as a function of temperature.}
   \label{fig:DLzero}
\end{figure*}

In Fig.~\ref{fig:DLzero} the zero momentum electric form factor is plotted for the various temperatures for the coarser and finer lattices. 
The figure also shows the differences of taking into account all the configurations (''no cuts'') and of using our selection procedure. 
As can be observed, the discontinuity in $D_L(0)$ at the critical temperature is enhanced when the separation of phases is performed. 
On the other hand, comparing the coarser and finer lattice results, it seems that the separation of $D_L(0)$ between the various sectors above $T_c$ is
reduced when approaching the continuum limit. 

In order to try to understand if the separation vanishes in the continuum limit, we have performed an additional simulation using a $90^3\times 10$ lattice and $\beta=6.212$ which has a $a = 0.07135$ fm and a $T = 276.6$ MeV.  We have checked that this simulation is in the deconfined phase.
The computed gluon form factors reproduce the pattern observed and already reported in Figs.~\ref{fig:DL2_T_54}, \ref{fig:DL2_T_72}, \ref{fig:DT2_T_54} and \ref{fig:DT2_T_72}.
In Tab.~\ref{Dzero_Tc} we compare the results for $D_L(0)$ and $D_T(0)$ for the simulations we have performed just above $T_c$. 
The results suggest that  the separation of the electrical gluon form factors observed between the various $Z_3$ sectors above $T_c$ is not a lattice artifact. 
From our simulation closer to the continuum one can claim a $|D_L(0; \theta = 0)  - D_L(0; \theta = \pm 2\pi/3)| \approx 180$ MeV$^{-2}$, where $\theta$ stands for
the phase of the Polyakov loop.

\begin{figure*}[t]
   \begin{center}
      \begin{subfigure}{\columnwidth}
       \includegraphics[scale=0.3,angle=-90]{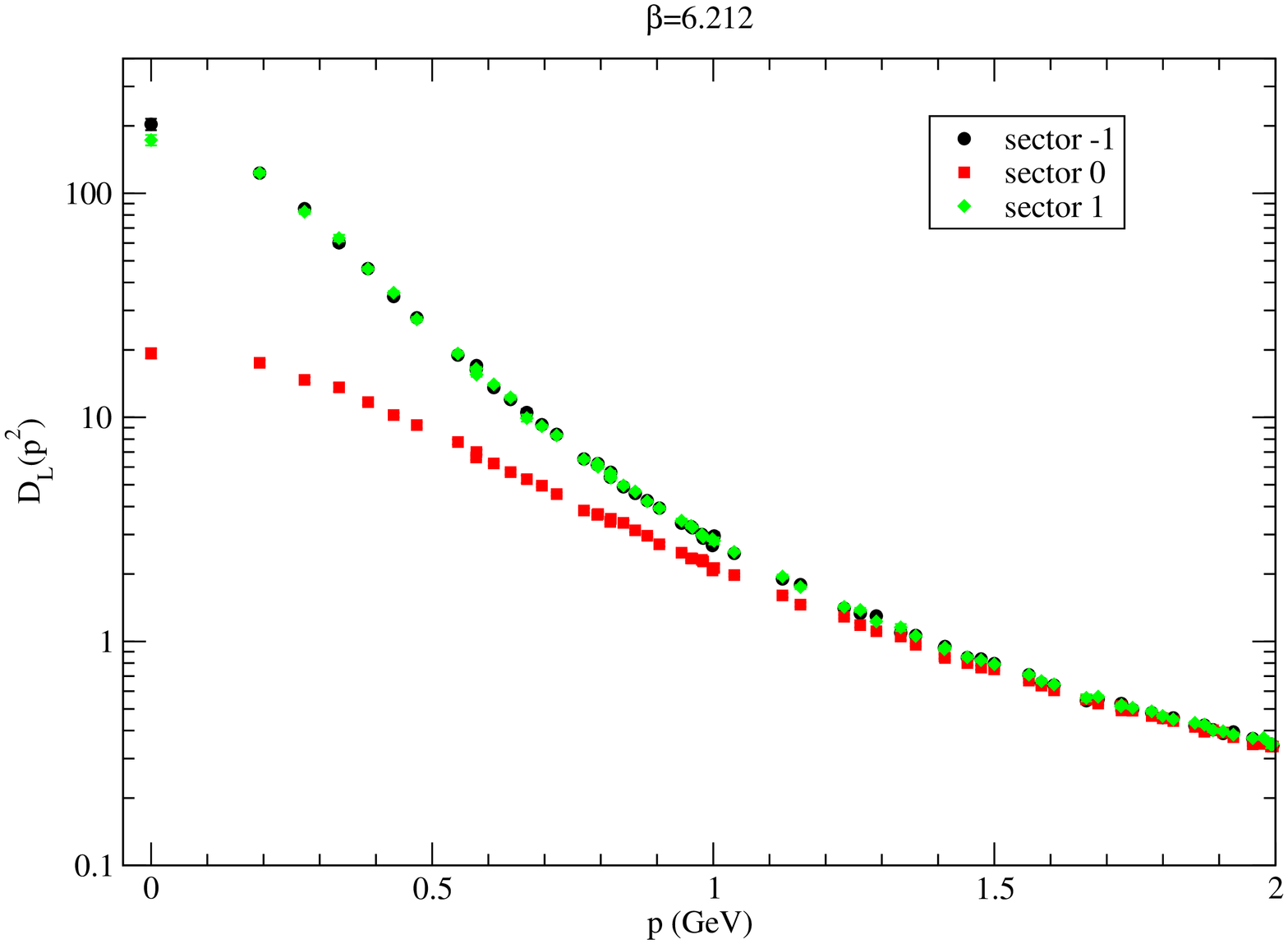} 
                        \caption{Longitudinal component.}
      \end{subfigure} \hfill
      \begin{subfigure}{\columnwidth}
      \includegraphics[scale=0.3,angle=-90]{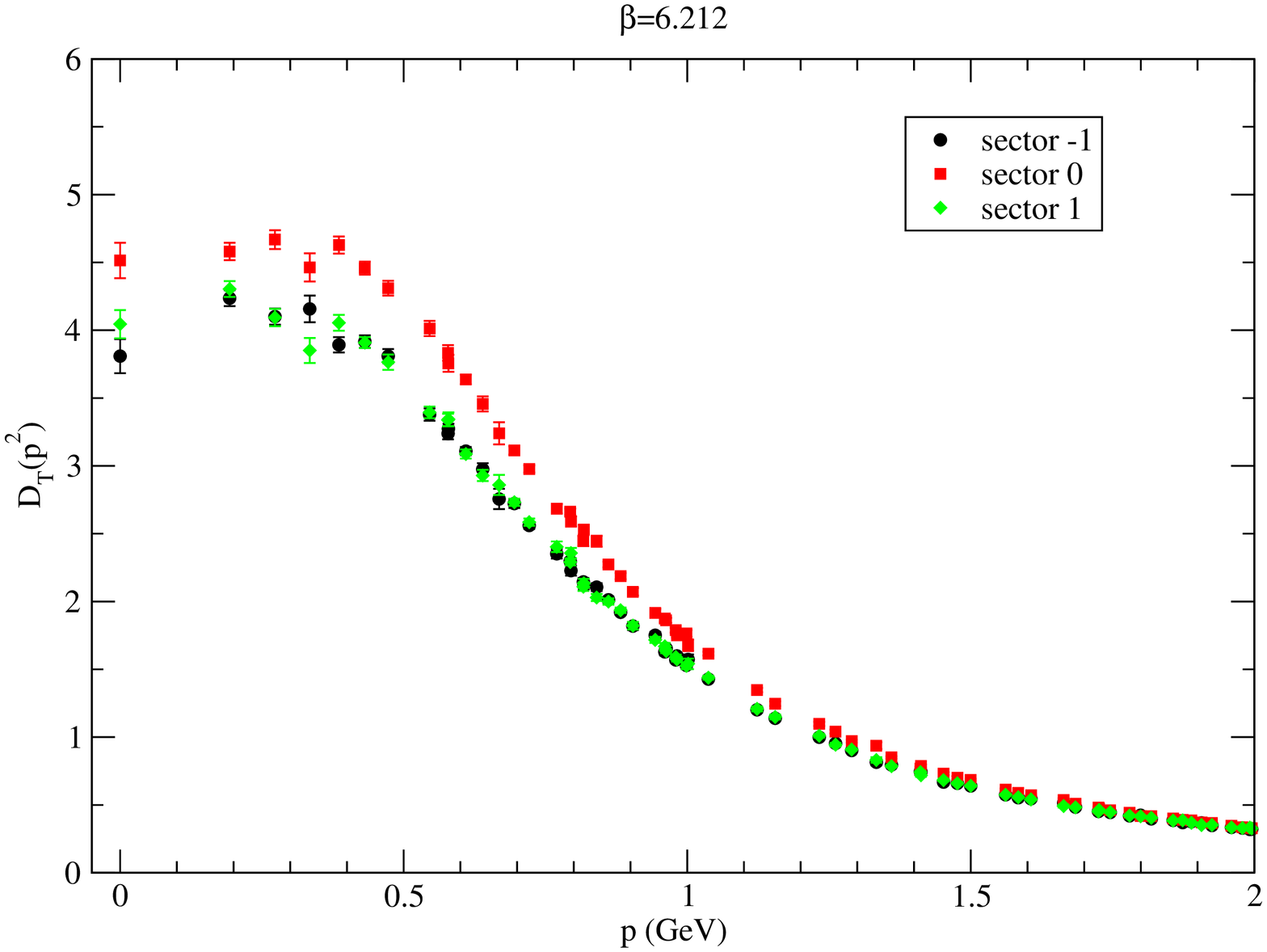} 
                       \caption{Transverse component.} 
                       \end{subfigure} 
  \end{center}
   \caption{Results for a simulation on a $90^3\times 10$ lattice, for $\beta=6.212$.  }
   \label{fig:D90}
\end{figure*}

%.......................................................................................................................................
%.......................................................................................................................................
\section{Summary and Conclusions \label{sec5}}

In this paper, we have investigated the correlations between the modulus of the Polyakov loop, its phase and the Landau gauge gluon propagator at finite temperature
for pure Yang-Mills SU(3) theory. In accordance with the literature, 
for temperatures above the deconfinement transition, the simulations show that the center symmetry is spontaneously broken and the
time history of the Monte Carlo reveals that the phase of the Polyakov loop is always close to $\theta \approx 0, \pm \, 2 \pi/3$. 
For temperatures below $T_c$, the Monte Carlo time history shows a $|L| \approx 0$.

We also discuss the computation of the gluon field and the gluon propagator for lattice configurations such that the phase of
the Polyakov loop is $\theta \neq 0$. 
 Our analysis shows that the usual definition given in Eq. (\ref{Eq:gluon_field0}) provides a valid
  way of computing the gluon field from the links and, therefore, the gluon propagator. 

For temperatures above $T_c$,  our simulations show that the gluon propagator associated to configurations with
$\theta \approx 0$ and $\theta \approx \pm \, 2 \pi/3$ differs quantitatively and qualitatively. For $T< T_c$, this difference is not observed. 
Therefore, the difference on
the propagators can be used to identify the phase, confined or deconfined, of a given configuration. Indeed,  this difference behaves as an 
order parameter for the confinement-deconfinement transition, vanishing for $T < T_c$ and taking finite non-zero values for $T > T_c$.

In what concerns the gluon propagator form factors above $T_c$,  it is observed a huge enhancement of the electric form factor and a suppression of the 
magnetic form factor for configurations where $\theta \approx \pm \, 2 \pi /3$ relative to those where $\theta \approx 0$. Once more, the simulations
show that it is in the electric sector where the dynamics is more sensitive to the deconfinement transition.

The pure Yang-Mills SU(3) theory has a first order transition to the deconfinement phase and the simulations performed for temperatures
near the critical temperature require a careful analysis.
Relying on the difference of the propagators associated to the various values of the phase of the Polyakov loop, we show that it is possible
to identify the phase, confined or deconfined, of a given configuration. Indeed, the criterion seems to be able to separate the configurations in each phase,
see Fig.~\ref{fig:D0L_T_all}, and this separation impacts directly on the computation of the propagator for temperatures near $T_c$, see Fig.~\ref{fig:DLzero}.
In fact, the effects of taking into account the configurations in either phase 
can be misunderstood as finite volume/size effects. For example, the ``systematic effects'' reported in  a recent analysis of the SU(2) gluon propagator close to the 
critical temperature~\cite{Mendes14} are possibly due to the mixing between the different nature of the gauge configurations generated by the Monte Carlo.
Note also that, as discussed in Sec.~\ref{sec4}, see Tab.~\ref{Dzero_Tc} and Fig.~\ref{fig:D90}, the observed difference
between the gluon form factors computed in different sectors seems to survive in the continuum limit.

The criterion to separate confined and deconfined phases based on the differences of the gluon propagator associated with the various $\theta$ values
also allows us to estimate the critical temperature $T_c $ in the range 269--272 MeV, in good agreement with the literature.

The gluon propagator is not a renormalization group  invariant quantity and the difference between the various sectors observed in 
the infrared region depends on the renormalization scale. In the current work, all the data was renormalized at $\mu = 4$ GeV and
the differences are clearly seen in the infrared region.
These differences in the propagators associated with the various $\theta$ are both quantitive and qualitative.
In principle,
 one could choose a different renormalization scale, e.g. in the infrared region, and, in this case, the differences in the propagators would appear in the ultraviolet
region and all the considerations discussed would still apply but for this region of momenta.

In a near future, we plan to extend our simulations to cover a wide range of temperatures to provide a clear picture on the behaviour of the various $D_L$, $D_T$ and
differences between the $Z_3$ sectors in a wide range of temperatures.
Furthermore, we are aware that the real world contains quark degrees of freedom and the center symmetry is no longer a valid symmetry of the theory. However, for full QCD 
the behaviour of the Polyakov loop as a function of the temperature is similar to the pure Yang-Mills case and, therefore, it would also be interesting to check how
the scenario discussed here changes when the quark degrees of freedom are taken into account.

\begin{acknowledgments}
The authors acknowledge the Laboratory for Advanced Computing at University of Coimbra for providing HPC computing resources (Milipeia, Centaurus and Navigator) that have contributed to the research results reported within this paper (URL http://www.lca.uc.pt). The authors also acknowledge computing resources provided by the Partnership for Advanced Computing in Europe (PRACE) initiative under DECI-12 project COIMBRALATT2. P. J. Silva ack\-now\-led\-ges support by F.C.T. under con\-tract SFRH/BPD/40998/\-2007. This work was sup\-por\-ted by projects CERN/FP/123612/2011, CERN/FP/123620/2011 and PTDC/FIS/100968/2008, projects developed under initiative QREN financed by UE/FEDER through Programme COMPETE.
\end{acknowledgments}

%%%%%%%%%%%%%%%%%%%%%%%%%   Bibliography   %%%%%%%%%%%

\end{document}